\documentclass[fleqn,usenatbib]{mnras}

\usepackage{newtxtext,newtxmath}

\usepackage[T1]{fontenc}

\newcommand{\msun}{\ensuremath{M_{\odot}}}
\newcommand{\lum}{erg\,s$^{-1}$}
\newcommand{\fermi}{{\it Fermi}}

\newcommand{\xmm}{{\it XMM-Newton}}
\newcommand{\swift}{{\it Swift}}
\newcommand{\chandra}{{\it Chandra}}

\newcommand{\ergflux}{\mbox{${\rm \, erg \,\, cm^{-2} \, s^{-1}}$}}
\newcommand{\gm}{$\gamma$}
\newcommand{\mbh}{$M_{\rm BH}$}

\newcommand{\km}{\mbox{${\rm \, km \,\,  s^{-1}}$}}

\newcommand{\bd}{{\tt BADASS}}

\newcommand{\hbeta}{H{$\beta$}}
\newcommand{\halpha}{H{$\alpha$}}
\newcommand{\FeII}{Fe{\sevenrm II}}

\newcommand{\OIII}{[O{\sevenrm\,III}]}
\newcommand{\OIIIa}{[O{\sevenrm\,III}]\,$\lambda$4959}
\newcommand{\OIIIb}{[O{\sevenrm\,III}]\,$\lambda$5007}

\newcommand{\NII}{[N{\sevenrm\,II}]}
\newcommand{\NIIa}{[N{\sevenrm\,II}]\,$\lambda$6549}
\newcommand{\NIIb}{[N{\sevenrm\,II}]\,$\lambda$6585}
\newcommand{\NIIab}{[N{\sevenrm\,II}]\,$\lambda\lambda$6549,6585}

\newcommand{\SIIa}{[S{\sevenrm\,II}]\,$\lambda$6718}
\newcommand{\SIIb}{[S{\sevenrm\,II}]\,$\lambda$6732}

 \font\sevenrm=cmr7 scaled 1000
\DeclareRobustCommand{\VAN}[3]{#2}
\let\VANthebibliography\thebibliography
\def\thebibliography{\DeclareRobustCommand{\VAN}[3]{##3}\VANthebibliography}

\usepackage{graphicx}	
\usepackage{amsmath}	

\title[NLSy1 galaxies in SDSS-DR17]{Narrow-line Seyfert 1 galaxies in Sloan Digital Sky Survey: a new optical spectroscopic catalogue}

\author[Paliya, Stalin, Dom{\'{\i}}nguez, Saikia]{
Vaidehi S. Paliya,$^{1}$\thanks{E-mail: vaidehi.s.paliya@gmail.com (VSP)}
C. S. Stalin,$^{2}$
Alberto Dom{\'{\i}}nguez,$^{3,4}$
and D. J. Saikia$^{1}$
\\
$^{1}$Inter-University Centre for Astronomy and Astrophysics (IUCAA), SPPU Campus, Pune 411007, India\\
$^{2}$Indian Institute of Astrophysics, Block II, Koramangala, Bengaluru 560034, Karnataka, India\\
$^{3}$Instituto de F\'isica de Part\'iculas y del Cosmos (IPARCOS), Universidad Complutense de Madrid, E-28040 Madrid, Spain\\
$^{4}$Departamento de Estructura de la Materia, F\'isica T\'ermica y Electr\'onica, Universidad Complutense de Madrid, E-28040 Madrid, Spain\\
}

\date{Accepted XXX. Received YYY; in original form ZZZ}

\pubyear{2023}

\begin{document}
\label{firstpage}
\pagerange{\pageref{firstpage}--\pageref{lastpage}}
\maketitle

\begin{abstract}
Narrow-line  Seyfert 1 (NLSy1) galaxies are an enigmatic class of active galactic nuclei (AGN) that exhibit peculiar multiwavelength properties across the electromagnetic spectrum. For example, these sources have allowed us to explore the innermost regions of the central engine of AGN using X-ray observations and have also provided clues about the origin of relativistic jets considering radio and \gm-ray bands.  Keeping in mind the ongoing and upcoming wide-field, multi-frequency sky surveys, we present a new catalogue of NLSy1 galaxies. This was done by carrying out a detailed decomposition of $>$2 million optical spectra of quasars and galaxies from the Sloan Digital Sky Survey Data Release 17 (SDSS-DR17) using the publicly available software ``Bayesian AGN Decomposition Analysis for SDSS Spectra". The catalogue contains 22656 NLSy1 galaxies which is more than twice the size of the previously identified NLSy1s based on SDSS-DR12. As a corollary, we also release a new catalogue of 52273 broad-line Seyfert 1 (BLSy1) galaxies. The estimated optical spectral parameters and derived quantities confirm the previously known finding of NLSy1 galaxies being AGN powered by highly accreting, low-mass black holes. We conclude that this enlarged sample of NLSy1 and BLSy1 galaxies will enable us to explore the low-luminosity end of the AGN population by effectively utilizing the sensitive, high-quality observations delivered by ongoing/upcoming wide-field sky surveys. The catalogue has been made public at \url{https://www.ucm.es/blazars/seyfert}.

\end{abstract}

\begin{keywords}
techniques:spectroscopic -- galaxies:active -- galaxies:Seyfert
\end{keywords}

\section{Introduction}\label{sec1}
Narrow-line Seyfert 1 (NLSy1) galaxies were, initially defined as, low-luminosity active galactic nuclei \citep[AGN, absolute $B$-band magnitude $M_{\rm B}>-23$,][]{1983ApJ...269..352S} which were identified purely based on their optical spectroscopic properties. \citet[][]{1985ApJ...297..166O} originally proposed to classify them based on the presence of narrow `broad permitted lines' and the strength of the forbidden \OIII~emission line compared to \hbeta~line with flux ratio of \OIIIb/\hbeta$<$3.  The spectropolarimetric study of NLSy1 galaxies was later carried out by \citet[][]{1989ApJ...342..224G} who quantified the first selection criterion of \citet[][]{1985ApJ...297..166O} with the full width at half maximum (FWHM) of the broad \hbeta~line to be $<$2000 \km. These objects exhibit several peculiar observational features such as the strong permitted \FeII~complexes, steep soft X-ray spectra, rapid X-ray flux variations \citep[cf.][]{1996A&A...305...53B,1999ApJS..125..297L,1999ApJS..125..317L} and strong outflows \citep[e.g.,][]{2002ApJ...565...78B,2008ApJ...680..926K,2010ApJS..187...64G,2012AJ....143...83X}. These observations have indicated the existence of rapidly accreting, low mass black hole systems (\mbh$\sim10^{6-8}$ \msun) powering these enigmatic AGN \citep[e.g.,][]{2000ApJ...542..161P,2004ApJ...606L..41G,2012AJ....143...83X}. However, alternative theoretical models have also been put forward by attributing the observed characteristics of NLSy1 galaxies to geometrical parameters, e.g.,  covering factor,  leading to the proposition that NLSy1 source population has been preferentially viewed at small angles compared to their broad line counterparts \citep[cf.][]{2008MNRAS.386L..15D}.

NLSy1 galaxies have been used to study a variety of AGN physics problems.  For example, the first fundamental correlation vector or the eigenvector 1 of Type 1 AGN (EV1) which represents the correlations of various observables such as the steep X-ray spectrum or strength of the \OIII~or optical \FeII~emission with the \hbeta~line width, has been argued to be an important Type 1 AGN unification scheme found so far \citep[][]{1992ApJS...80..109B,2000ARA&A..38..521S}.  These objects have been found to lie at the extreme negative end of EV1 thereby providing us clues about the central engine parameters, e.g., black hole mass or accretion rate, and/or geometrical aspects such as the viewing angle \citep[cf.][]{1996A&A...305...53B,2000ARA&A..38..521S,2001ApJ...558..553M}. Furthermore, the X-ray spectra of NLSy1 galaxies often show a strong soft X-ray excess and reflection dominated hard X-ray emission \citep[e.g.,][]{2009Natur.459..540F}. Deep X-ray observations of these sources have permitted us to study the behaviour of matter and energy and their possible interaction in the immediate vicinity of the central supermassive black hole \citep[][]{2014MNRAS.443.1723P,2017MNRAS.468.3489K}. Though most of the NLSy1 galaxies are radio-quiet, $\sim$5\% of them are found to be radio-loud indicating the presence of jets \citep[][]{2006AJ....132..531K,2008ApJ...685..801Y}. Some of the very radio-loud NLSy1s have also been detected in the all-sky \gm-ray survey being conducted with the \fermi-Large Area Telescope \citep[][]{2009ApJ...707L.142A,2018ApJ...853L...2P}. This has led to the idea of them being the nascent blazars \citep[e.g.,][]{2020ApJ...892..133P}. Indeed, the general NLSy1 population has been considered as rapidly accreting, low-luminosity AGN in the early stage of their evolution \citep[cf.][]{2000MNRAS.314L..17M,2018A&A...614A..87B}. Moreover, these sources have also been reported to exhibit rapidly rising and long-lasting optical flaring activity thereby making them crucial targets in the era of time-domain astronomy \citep[][]{2021ApJ...920...56F}.

The above-mentioned research problems highlight the pivotal role that NLSy1 galaxies can play in advancing our current understanding of AGN science.  Also considering the latest and upcoming wide-field, multi-frequency sky surveys and missions, e.g., Very Large Array Sky Survey \citep[VLASS,][]{2020PASP..132c5001L}, eROSITA \citep[][]{2021A&A...647A...1P}, and Rubin observatory \citep[][]{2019ApJ...873..111I}, it is imperative to increase the sample size of the known NLSy1 galaxies. The latest catalogue of this class of AGN was prepared using the Sloan Digital Sky Survey data release 12 (SDSS-DR12) and contains 11,101 NLSy1s \citep[][]{2017ApJS..229...39R} which superseded the earlier catalogues containing 150 and 2011 sources using SDSS early data release and SDSS-DR3, respectively \citep[][]{2002AJ....124.3042W,2006ApJS..166..128Z}. Since then, there have been a number of major updates, both in data collection and analysis software,  that motivated us to prepare a new sample of these enigmatic AGN. We highlight a few such updates below:

\begin{enumerate}
\item The number of spectroscopically observed sources has considerably increased in the most recent DR17 which is the final survey from the fourth phase of SDSS \citep[][]{2017AJ....154...28B}.  For example, SDSS-DR16 contains 750,414 quasars\footnote{SDSS-DR-17 does not contain any new spectroscopically observed quasars.} \citep[][]{2020ApJS..250....8L} which is more than two times larger than the 297,301 objects published in DR12 \citep[][]{2017A&A...597A..79P}.
\item The work of \citet[][]{2017ApJS..229...39R} only considered objects classified as quasars (SDSS pipeline keyword {\tt QSO}). In addition to quasars, we have searched for NLSy1s among $\sim$1.9 million sources identified as galaxies by the SDSS data reduction pipeline (keyword {\tt GALAXIES}).
\item In \citet[][]{2017ApJS..229...39R}, the analysis of SDSS spectra was done using a custom-built private software. In this work, we have instead adopted the publicly available optical spectroscopic data analysis software `Bayesian AGN Decomposition Analysis for SDSS Spectra\footnote{\url{https://github.com/remingtonsexton/BADASS3}}' \citep[\bd,][]{2021MNRAS.500.2871S}. This package is an open-source spectral analysis tool designed for detailed decomposition of SDSS spectra.
\item Analyzing SDSS spectra with the motivation to identify NLSy1 galaxies will naturally lead to finding new broad-line Seyfert 1 (BLSy1) galaxies. While earlier studies have only reported the NLSy1 galaxy sample \citep[][]{2002AJ....124.3042W,2006ApJS..166..128Z,2017ApJS..229...39R}, we release the catalogues of both NLSy1 and BLSy1 sources.

\end{enumerate}

In Section~\ref{sec2}, we briefly describe the sample selection criteria, while details of the \bd~spectroscopic data analysis steps are elaborated in Section~\ref{sec3}. The new catalog of NLSy1 galaxies and their multi-wavelength properties are discussed in Sections~\ref{sec4} and~\ref{sec5}, respectively. We summarize our findings in Section~\ref{sec6}. Throughout, a flat cosmology with $H_0 = 70~{\rm km~s^{-1}~Mpc^{-1}}$ and $\Omega_{\rm M} = 0.3$ was adopted.

\section{Sample Selection}\label{sec2}
We considered all DR-17 sources that were classified either as {\tt QSO} (1370779 objects) or {\tt GALAXIES} (3237535 sources) by the SDSS pipeline.  Next, we applied the following filters to retain sources for spectroscopic analysis:
\begin{enumerate}
\item reject all sources with $z>$0.8 and $>$0.9 for SDSS and Baryon Oscillation Spectroscopic Survey (BOSS) spectrographs, respectively. This is because the optical spectra taken with the BOSS spectrographs have larger wavelength coverage compared to the SDSS spectrographs. This redshift cut ensured that both \hbeta~and \OIII~emission lines, needed to characterize a NLSy1 galaxy, are present in the optical spectrum. 
\item reject all objects with the keyword {\tt zWarning} $>$0 and {\tt zWarning} $\neq 16$. The latter condition, i.e.,  {\tt zWarning} $=16$, usually indicates a high S/N spectrum or broad emission lines in a galaxy\footnote{\url{https://www.sdss4.org/dr17/algorithms/bitmasks/\#ZWARNING}}.
\item reject all sources with the keyword {\tt specprimary}$=0$ since the best observations of all unique objects have {\tt specprimary} $>$0.
\item reject all objects with median S/N ratio {\tt snMedianAll} $<$2.
\end{enumerate}

This exercise left us with 111506 {\tt QSO} and 1918262 {\tt GALAXIES}. Furthermore, we also included 11754 quasars from the SDSS-DR16 quasar catalog that were left out possibly due to one/more filters mentioned above.  Recently, \citet[][]{2022ApJS..263...42W} reported the results of the optical spectroscopic analysis of SDSS-DR16 quasars and also provided improved redshifts measurements. We used their redshift measurements for sources common in both samples and adopted the SDSS pipeline redshift for the rest of them. Overall, our final sample contains 123260 quasars and 1918262 galaxies. The analyses of these $>$2 million spectra were carried out using \bd~software as described in the next section.

\section{Optical spectroscopic data analysis}\label{sec3}
\subsection{Bayesian AGN Decomposition Analysis for SDSS Spectra}

\begin{figure*}
\vbox{
\includegraphics[scale=0.6]{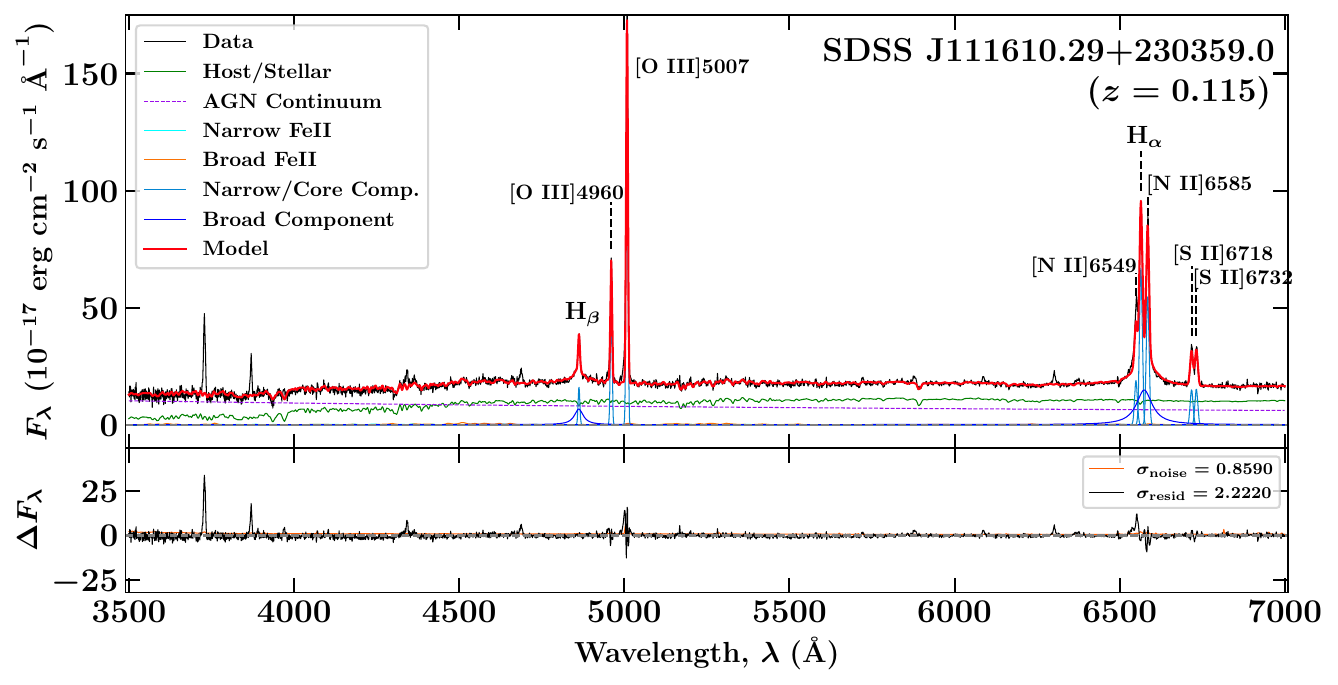}
\includegraphics[scale=0.6]{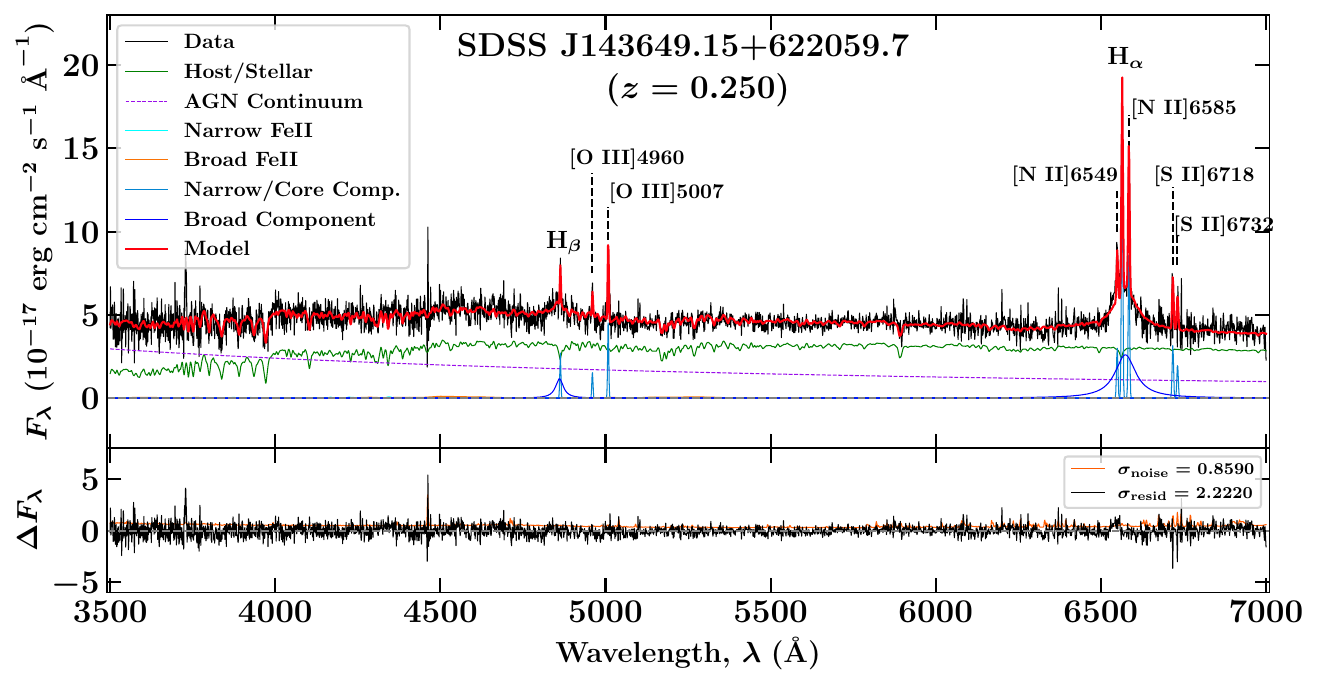}
}
\caption{Examples of the spectral fitting of a quasar (top) and a galaxy (bottom) spectra released in SDSS-DR17. The modeled components are labeled.} \label{fig:fit}
\end{figure*}

\bd~is an open source optical spectroscopic data analysis package designed to automate the deconvolution of AGN and host galaxy spectra,  simultaneously fitting both emission line and continuum features, and estimating robust parameter uncertainties using Markov Chain Monte Carlo (MCMC) approach. The full description of the tool can be found in \citet[][]{2021MNRAS.500.2871S} and \citet[][]{2022ApJS..260...33S} and here we summarize its salient features.

In the first step, the optical spectrum is brought to the rest-frame and corrected for the Galactic extinction using the extinction map of \citet[][]{2011ApJ...737..103S} and extinction law of \citet[][]{1989ApJ...345..245C} considering $R_{\rm V}=3.1$. Next, the software simultaneously fits all components, e.g., emission lines,  to constrain their relative contribution and covariances.

To model the continuum emission, \bd~uses a composite host-galaxy template \citep[e.g.,][]{2016MNRAS.463.3409V} or empirical stellar templates to estimate the line-of-sight velocity distribution \citep[LOSVD, cf.][]{2017MNRAS.466..798C}, a power-law and Balmer pseudocontinuum components mimicking the AGN contribution \citep[cf.][]{2002ApJ...564..581D}, and optical-UV \FeII~emission \citep[][]{2001ApJS..134....1V,2004A&A...417..515V}.

The \bd~software provides several line profile shapes, e.g., Gaussian or Lorentzian, to fit the broad and narrow emission lines. The measured line widths are corrected for the SDSS instrumental resolution. The flux ratios of \OIII~and \NII~doublets are kept fixed to 3 during the fit \citep[e.g.,][]{2011ApJS..194...45S}. Optionally, the widths of narrow lines can also be tied, e.g., all narrow line width components in the \hbeta~region are tied to the \OIII~line width.

The model fitting in \bd~is done using the Bayesian affine-invariant MCMC sampler {\tt emcee} \citep[][]{emcee}. To obtain the initial parameter values, a maximum-likelihood fit is performed adopting a Gaussian likelihood distribution. The full parameter space is then scanned using MCMC to derive robust parameter and uncertainty estimates. Additionally, the software also provides an option of performing multiple iterations of maximum-likelihood fitting by applying a Monte Carlo bootstrapping technique.  The spectra are perturbed by adding a random normally distributed noise at every pixel using the spectral flux uncertainties and re-fitting the spectra. The median (50th percentile) of the distribution is adopted as the best-fit values of the spectral parameters and the semi amplitude of the range covering the 16th-84th percentiles of the distributions is considered as 1$\sigma$ uncertainties on each parameter.

\subsection{Spectral Analysis of Quasars}
The spectral fitting of 123260 SDSS quasars was done in the wavelength range 3500$-$7000 \AA. The host galaxy emission was modeled with three single stellar population templates of ages 0.1, 1, and 10 Gyr from EMILES library \citep[][]{2016MNRAS.463.3409V}. We did not attempt applying more sophisticated penalized Pixel-Fitting software \citep[{\tt pPXF},][]{2017MNRAS.466..798C} to model the host galaxy emission since the overall spectrum is expected to be dominated by the emission lines.  To reproduce the continuum, we also considered a power-law and optical \FeII~complex template from \citet[][]{2004A&A...417..515V}. Since Balmer lines in NLSy1 galaxies are usually better represented by a Lorentzian function \citep[cf.][]{2002ApJ...566L..71S,2012MNRAS.426.3086G,2016MNRAS.462.1256C}, we fitted the broad components of \halpha~and \hbeta~emission lines with a Lorentzian function.  The narrow emission line profiles, on the other hand, were modeled with a Gaussian function. The \OIII~doublet was fitted with a single or double Gaussian functions, one each for the core and wing, depending on the line shapes and signal-to-noise (S/N) ratio of the data. We also applied the condition that the width of the broad components must be larger than that of the narrow emission lines.  A maximum-likelihood fitting technique using a normal likelihood distribution was adopted to carry out the initial optimization of the spectral parameters which was followed by MCMC fitting for the robust parameter and uncertainty estimations. A maximum of 5000 iterations of MCMC sampling were performed with 100 walkers per parameter and we considered the final 1000 iterations (4000 burn-in) for the posterior distributions. The median of the distribution was taken as the best-fit value and the 16th and 84th percentiles of the distributions were used as the lower- and upper-bound 1$\sigma$ uncertainties on each measured quantity. Figure~\ref{fig:fit} shows the fitted optical spectrum of one of the analyzed quasars.

\subsection{Spectral Analysis of Galaxies}
The optical spectral analysis of $\sim$1.9 million SDSS galaxies was carried out following a strategy similar to that adopted to model the quasar spectra. However, given the large number of sources and to optimize the computational resources, we divided the analysis in two parts. First, we modeled the continuum with a power law and optical \FeII~template and also applied the {\tt pPXF} software to reproduce the strong host galaxy emission. The emission lines were modeled with Gaussian functions and the S/N ratio of the broad \hbeta~emission line was computed. To speed up the process, results obtained from a single maximum-likelihood fitting were considered.  At this stage, we rejected all sources in which the S/N ratio of the broad \hbeta~emission line was found to be $<$2. This exercise led to the selection of 1143 galaxies. Then, we repeated the full spectral fitting, similar to that adopted for quasars, on these objects and estimated the parameters and uncertainties by applying the MCMC fitting technique. An example of the fitting is demonstrated in Figure~\ref{fig:fit}.

\begin{figure}
\includegraphics[width=\columnwidth]{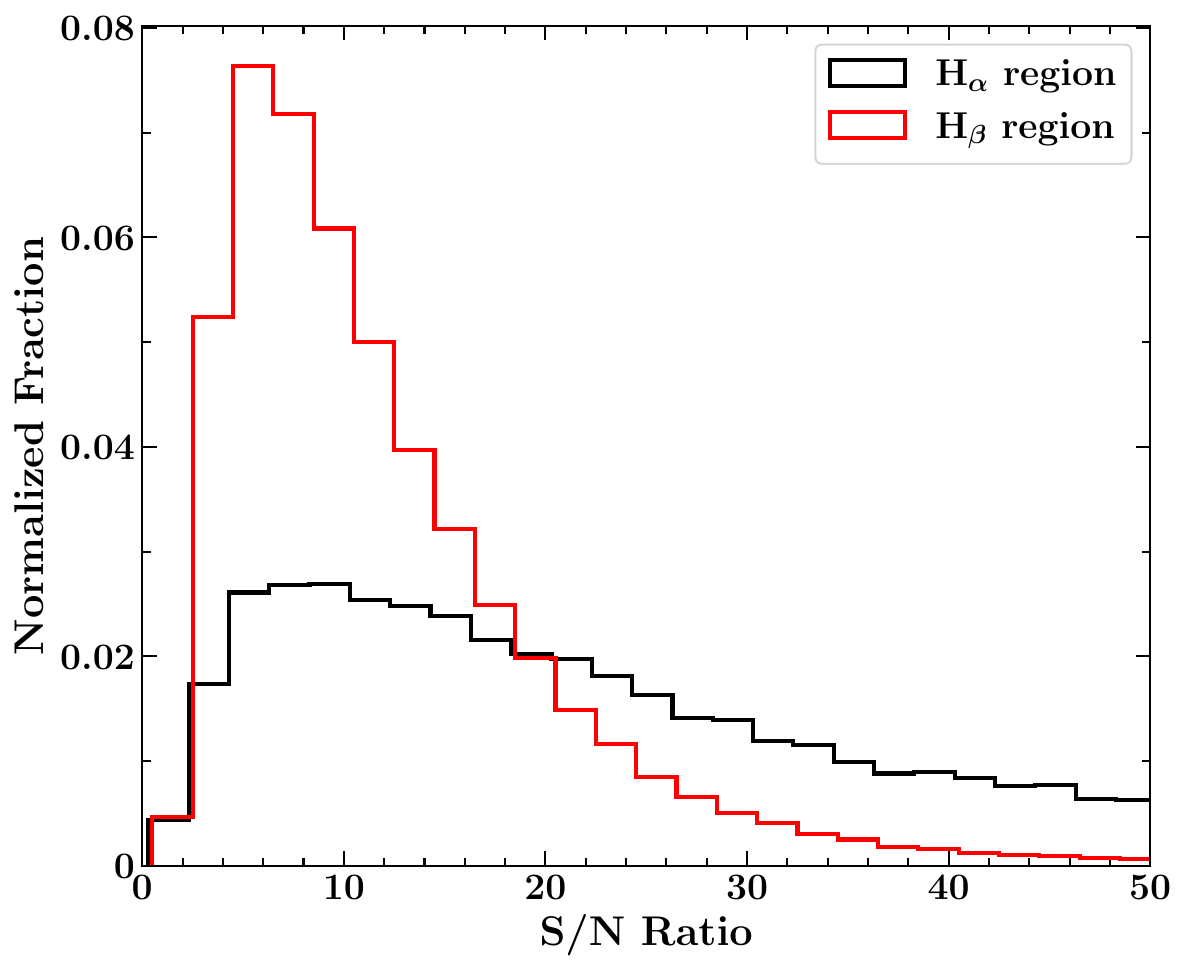}
\caption{This plot shows the distributions of the measured S/N ratio for the \halpha~and \hbeta~regions. The areas of the plotted histograms are normalized to unity.} \label{fig:snr}
\end{figure}
\subsection{Reliability of Spectral Modeling}
The \bd~software carries out the spectral data analysis in fully automatic mode.  As usual for any modeling technique, the reliability of the fitting results strongly depends on the S/N ratio of the spectrum. In Figure~\ref{fig:snr}, we show the distribution of the S/N ratio of the \hbeta~and \halpha~emission line regions.  For \hbeta~line, majority of sources have S/N ratio $\sim$5 possibly due to the fact that DR17 quasars probe fainter flux limits compared to previous data releases given the improved sensitivity of BOSS spectrographs \citep[see, e.g., ][]{2011ApJS..194...45S,2022ApJS..263...42W}.  In this regard, the application of MCMC fitting that has allowed us to derive the robust parameter uncertainties by properly taking into account the data quality, ensures that the measured parameters are reliable. Unlike the conventionally adopted least-square fitting method which may get stuck to a local minima, MCMC fitting scans the full parameter space so that the global minima is achieved.  Furthermore, we also performed a number of checks and cross-matched our results with other published works to verify the robustness of the obtained results as described in the next section. We also provide the S/N ratio measured for all of the broad and narrow emission lines in our NLSy1 and BLSy1 catalogues so that a user can customize the selection of objects for a particular research problem. 

\section{Results}\label{sec4}
We derived the emission line parameters for 123260 quasars and 1143 galaxies using the \bd~software. All fitted spectra were visually inspected to filter out contaminating objects whose spectral parameters could not be well constrained. If required, the fit was repeated, e.g., with modified wavelength coverage to avoid noisy edges of the spectrum, and parameters were calculated again. This led to the rejection of 49428 optical spectra. A major fraction of these objects turned out to be either Type 2 AGN with no broad lines detected in their optical spectra or those with extremely poor quality data.  The spectral parameters of the remaining 74975 AGN, including 222 galaxies, were further analyzed to identify NLSy1 and BLSy1 galaxies as discussed in the next section.

\begin{figure*}
\vbox{
\includegraphics[scale=0.31]{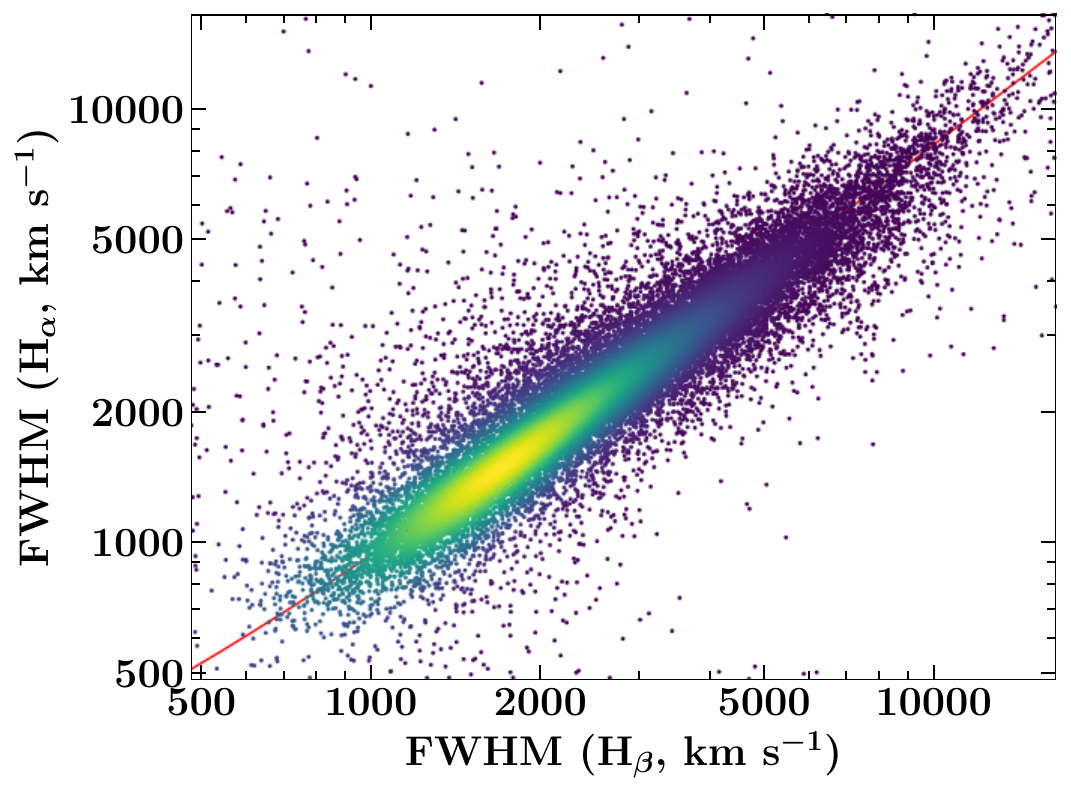}
\includegraphics[scale=0.31]{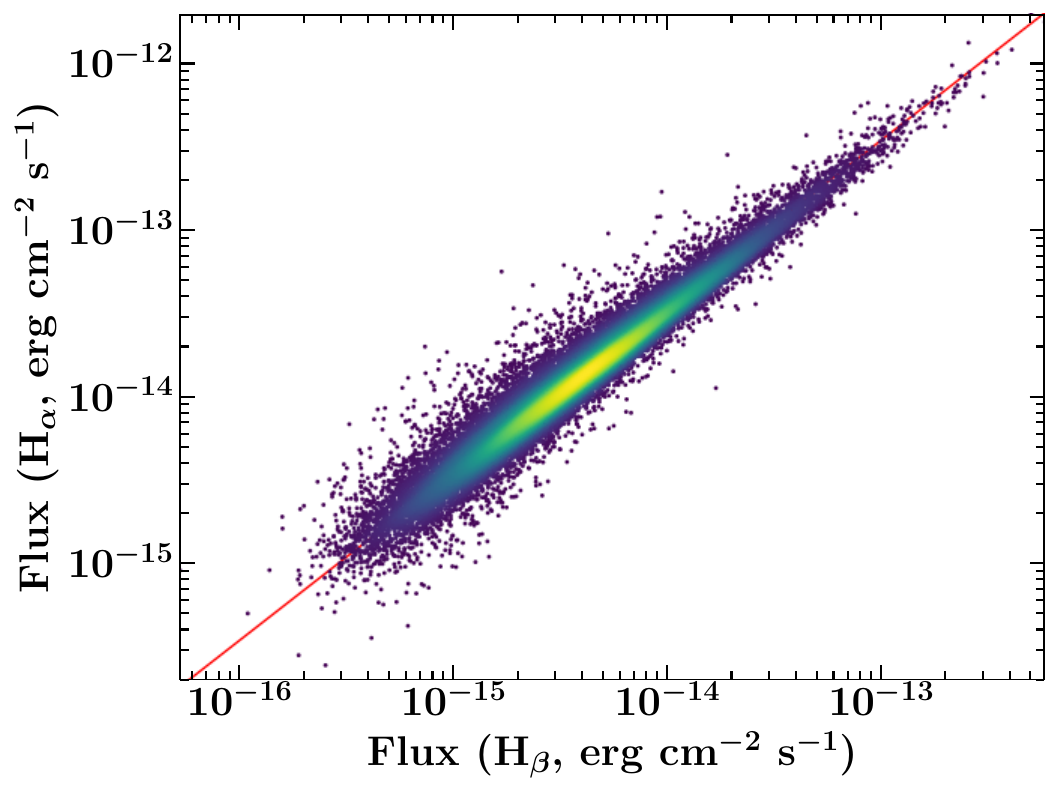}
\includegraphics[scale=0.31]{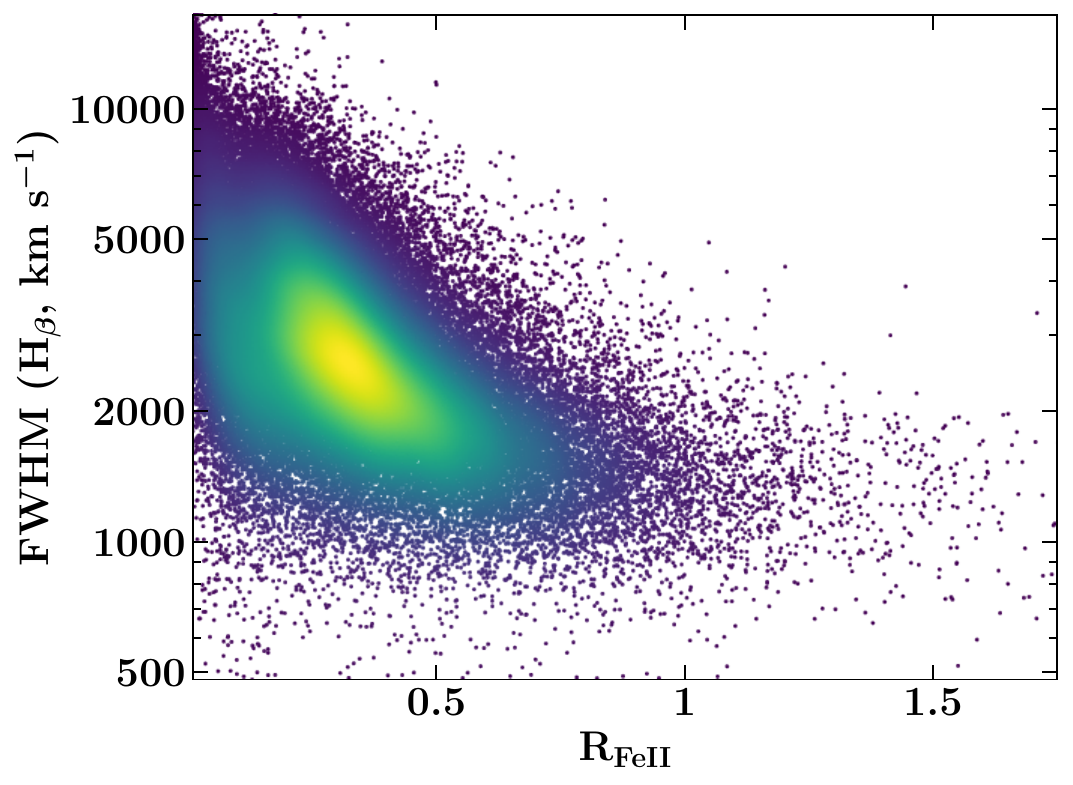}
}
\caption{The left and middle panels show the variation of the broad \halpha~emission line FWHM and flux, respectively, with that measured for the broad \hbeta~line. The red line refers to the best-fitted correlation for the plotted quantities. The distribution of the sources on the optical \FeII~strength versus \hbeta~line FWHM plane is shown in the right panel. In all plots, the number density of the data points is colour coded with lighter colours representing larger density of sources.} \label{fig:self}
\end{figure*}

In Figure~\ref{fig:self}, we plot some of the measured spectral parameters, namely FWHM and flux of the broad \halpha~and \hbeta~components for the full sample. We also show the variation of the FWHM of the broad \hbeta~line as a function of the optical \FeII~strength ($R_{\rm 4570}$) which is defined as the ratio of the \FeII~flux in the wavelength range 4434 \AA$-$4684 \AA~to the flux of the broad component of the \hbeta~emission line \citep[e.g.,][]{1992ApJS...80..109B}.  The FWHM of the broad components of the Balmer lines are found to be correlated as $FWHM_{\rm H\alpha}\propto(0.82\pm0.01)FWHM_{\rm H\beta}$ confirming the results found in earlier studies \citep[cf.][]{2006ApJS..166..128Z}. Similarly, the flux values of the broad \halpha~and \hbeta~lines are also correlated with $F_{\rm H\alpha}\propto(3.41\pm0.01)F_{\rm H\beta}$ \citep[e.g.,][]{2013ApJ...763..145D}.

In Appendix, Table~\ref{tab:nlsy1_cat} reports the measured spectral parameters for the sample which is published in the form of online NLSy1 and BLSy1 catalogues\footnote{\url{https://www.ucm.es/blazars/seyfert}}.  Along with the measured quantities, we also derived $B$-band absolute magnitude ($M_{\rm B}$), bolometric luminosity ($L_{\rm bol}$), single epoch virial black hole mass ($M_{\rm SE}$), and Eddington ratio ($R_{\rm Edd}$). 

We used the SDSS $g$ and $r$-filter magnitudes and adopted the following transformation equation to estimate the Bessel $B$-band magnitude \citep[][]{2006A&A...460..339J}:
\begin{equation}
B = g + (0.313 \pm 0.003) * (g - r) + (0.219 \pm 0.002),
\end{equation}
which was then used to estimate $M_{\rm B}$. The objective of calculating this parameter was to consider the fact that most of the studies on NLSy1 galaxies do not differentiate them with narrow-line quasars. Though we also do not apply any selection filter using the absolute magnitude based quasar/Seyfert classification \citep[$M_{\rm B}\lessgtr-23$,][]{1983ApJ...269..352S},  this piece of information may enable the user to select genuine NLSy1 objects. For the sake of consistency, we throughout mention all broad/narrow line objects as BLSy1/NLSy1 galaxies.

 We considered the commonly adopted single epoch virial technique to estimate the mass of the central black hole that assumes the broad line region to be virialized \citep[e.g.,][]{2006ApJ...641..689V}. In particular, the following equation was used to calculate $M_{\rm SE}$ \citep[e.g.,][]{2004MNRAS.352.1390M}:

\begin{equation}
M_{\rm SE} = \frac{\zeta R_{\rm BLR} V^2_{\rm BLR}}{G}
\end{equation}

where $V_{\rm BLR}$ is the Keplerian velocity of the line emitting BLR clouds, $R_{\rm BLR}$ is the BLR radius, and $G$ is the gravitational constant. The parameter $\zeta$ refers to the scale factor considering the kinematics and geometry of the BLR and taken as $\zeta=3/4$ assuming spherical distribution of clouds \citep[][]{2017ApJS..229...39R}. The $R_{\rm BLR}$ was estimated following \citet[][]{2019ApJ...886...42D} who updated the scaling relation between $R_{\rm BLR}$ and 5100 \AA~continuum luminosity including the relative strength of \FeII~emission, which is important for highly accreting AGN such as NLSy1 galaxies. In particular, the following equation was adopted:

\begin{equation}
\log\left(\frac{R_{\rm BLR}}{\rm lt-day}\right) = A + B\log\left({\lambda L_{\lambda,5100} \over 10^{44}\,{\rm erg\,s^{-1}}}\right) + C R_{\rm 4570}
\end{equation}

where coefficients $A$, $B$, and $C$ are estimated as 1.65, 0.45, and $-$0.35 \citep[][]{2019ApJ...886...42D}. The reported uncertainty in $M_{\rm SE}$ refers to measurement errors and does not take into account any systematics which can be as large as 0.4 dex \citep[e.g.,][]{2013BASI...41...61S}. Furthermore, we computed $L_{\rm bol}$ from 5100 \AA~continuum luminosity adopting the bolometric correction factor of 9.26  \citep[][]{2006ApJS..166..470R}. The Eddington ratio, $R_{\rm Edd}$, was estimated from the derived $M_{\rm SE}$ and $L_{\rm bol}$.

\subsection{The NLSy1 Catalogue}
To identify the genuine NLSy1 galaxies, we used the following two conditions:
\begin{enumerate}
\item the FWHM of the broad \hbeta~emission line within measured uncertainties is smaller than 2000 \km, i.e., $FWHM-\Delta FWHM \leq2000$ \km.
\item the flux ratio of \OIIIb~and \hbeta~emission lines within estimated uncertainties is $<$3, where we propagated the uncertainties in the \hbeta~and \OIIIb~flux values while computing the ratio.
\end{enumerate}

The first filter follows the classic definition of NLSy1 galaxies \citep[][]{1989ApJ...342..224G}. It is slightly different from that used in previous NLSy1 catalogues \citep[cf.][]{2006ApJS..166..128Z,2017ApJS..229...39R}.  These works used the \hbeta~FWHM threshold of 2200 \km~without considering the uncertainties in the measured quantity which we have accounted for. The second condition was proposed to separate NLSy1s from Seyfert 2  galaxies. To calculate the ratio,  the fluxes of those \OIIIb~and narrow \hbeta~emission lines were considered that have S/N ratio $>$1 and non-zero flux uncertainties, otherwise flux values were assumed zero since in such cases the line detection is marginal at best. The application of above two filters led to the final sample of 22656 NLSy1 galaxies present in SDSS-DR17. There are 46 sources that qualified the first selection filter but not the second, i.e., the flux ratio of \OIIIb~and \hbeta~emission lines was found to be $>$3.  If the threshold of FWHM of the broad \hbeta~component is relaxed to 2200 \km, the NLSy1 sample size grows to 27298. Nevertheless, we stick to the original threshold of 2000 \km~and list all other objects in the BLSy1 catalogue so that if a user wishes to use the relaxed criterion, they can find all of the information in this catalogue. Furthermore, 17206 sources have $M_{\rm B}>-23$ indicating a large fraction of the NLSy1 sample to be genuine Seyfert galaxies \citep[][]{1983ApJ...269..352S}. 

\subsection{Comparison with previous works}
\begin{figure}
\includegraphics[width=\linewidth]{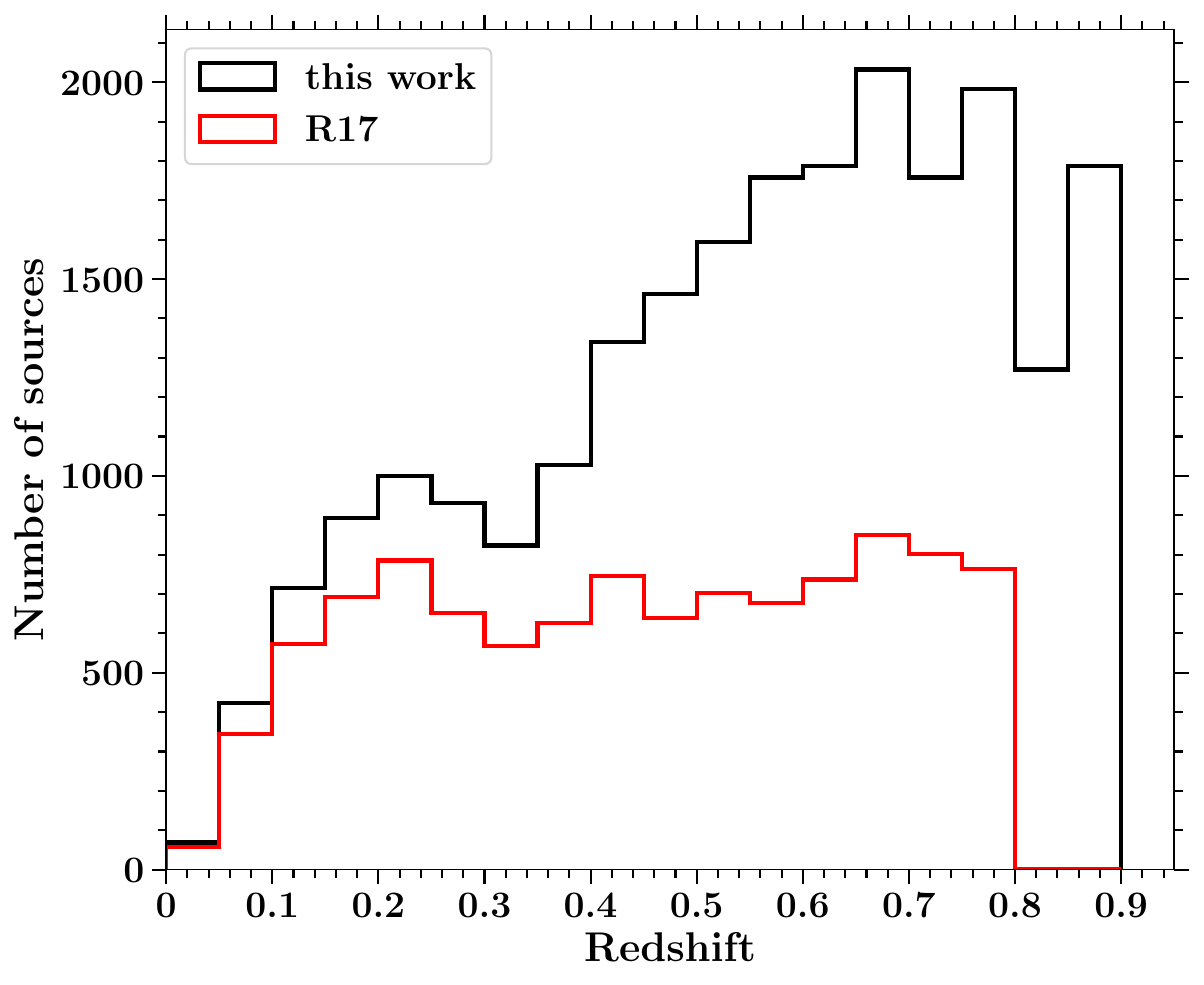}
\caption{The redshift histograms of NLSy1 galaxies present in our sample and that included in the SDSS-DR12 NLSy1 catalogue.} \label{fig:z_dis}
\end{figure}

For a consistency check, we compared our catalogue of NLSy1 galaxies with earlier published works. Cross-matching with the SDSS-DR12 NLSy1 catalogue \citep[][]{2017ApJS..229...39R} by using a maximum search radius of 3 arcsec, we found that 884 quasars classified as NLSy1 galaxies in SDSS-DR12 were rejected during the visual inspection.  Among the remaining 10217 objects, 8904 i.e., $\sim$87.1\%,  sources are common in both the catalogues. If we adopt the relaxed threshold of $FWHM_{\rm H\beta}=2200$ \km~to select NLSy1s, the match percentage increases to 94.5\% with 9650 sources present in both the works.  A comparison of the redshift distributions indicates that a majority of the newly identified NLSy1 galaxies in our work are at higher redshifts (Figure~\ref{fig:z_dis}).

\begin{figure*}
\hbox{\hspace{0cm}
\includegraphics[scale=0.3]{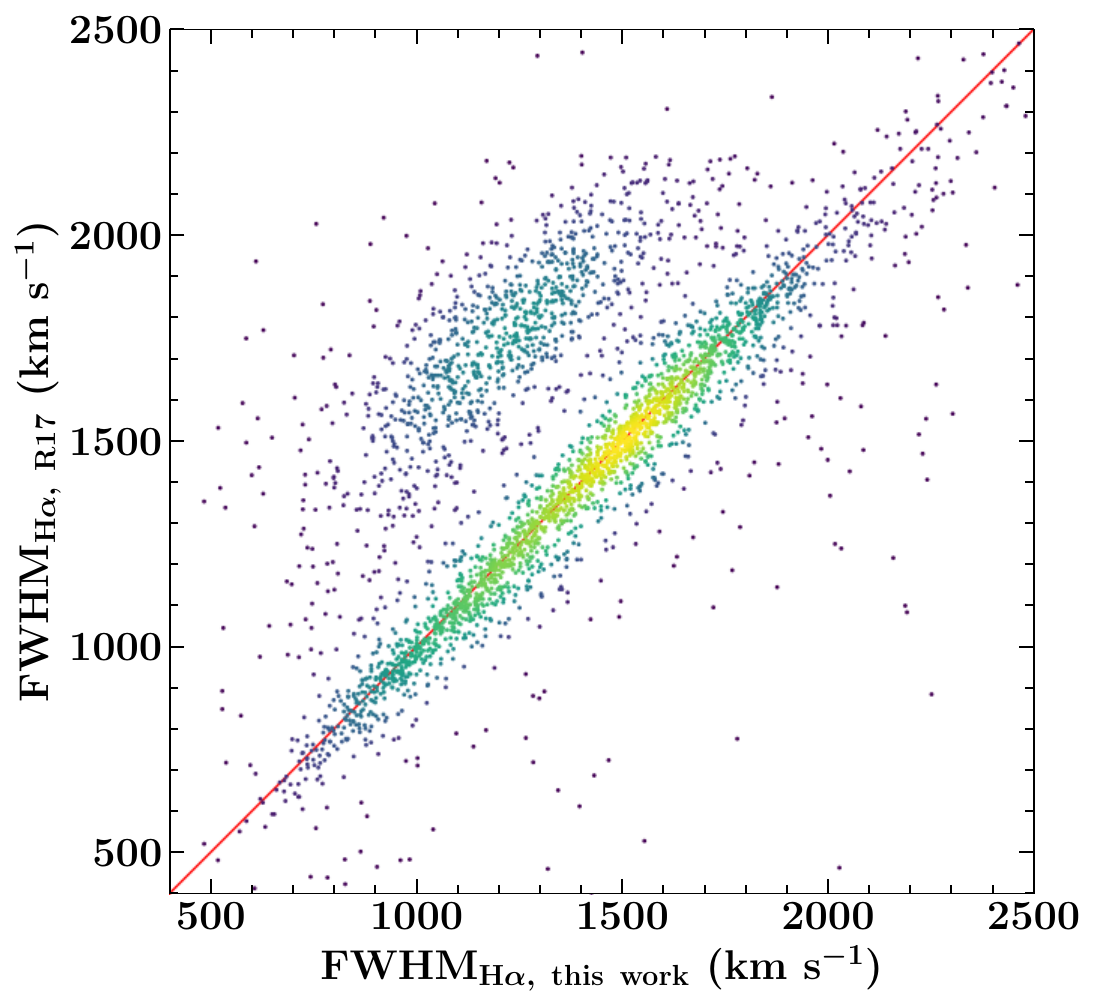}
\includegraphics[scale=0.3]{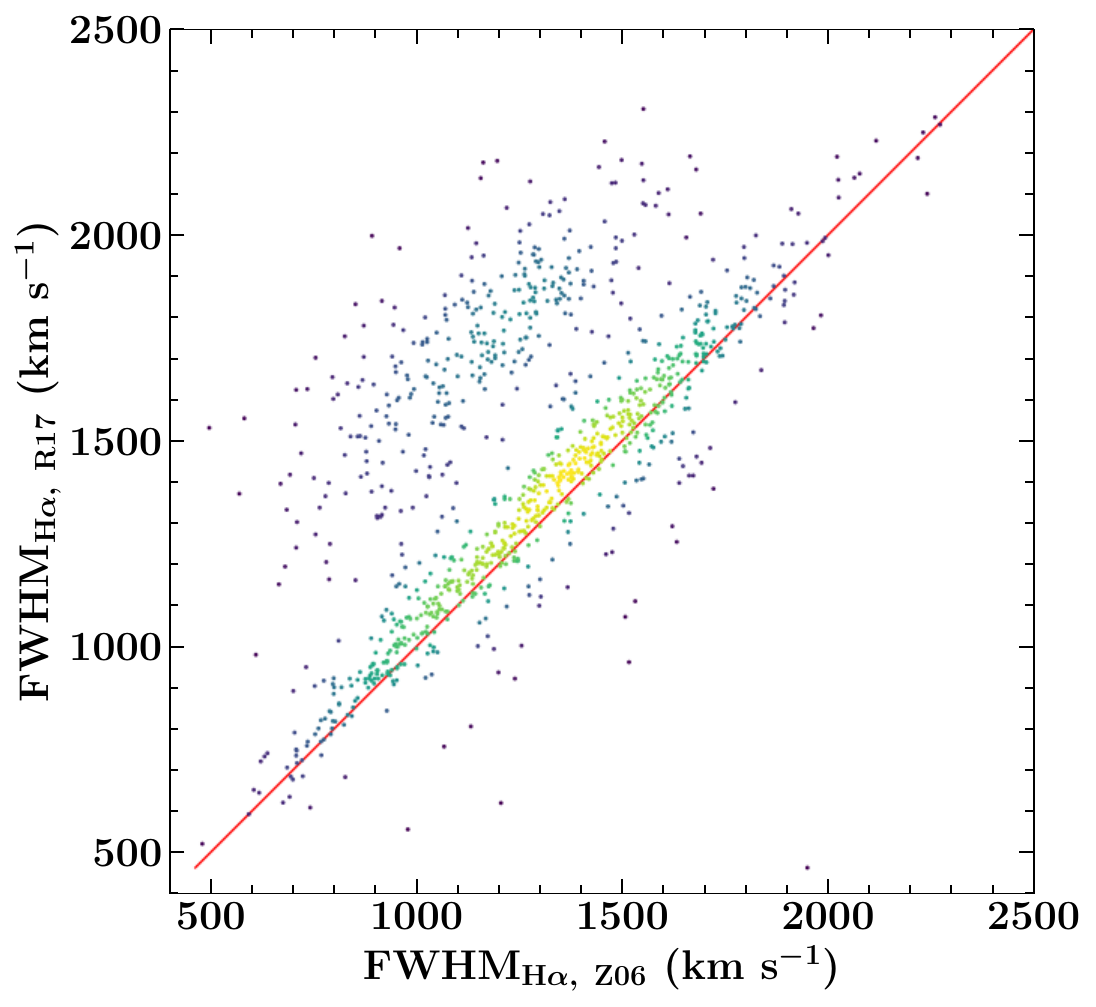}
\includegraphics[scale=0.3]{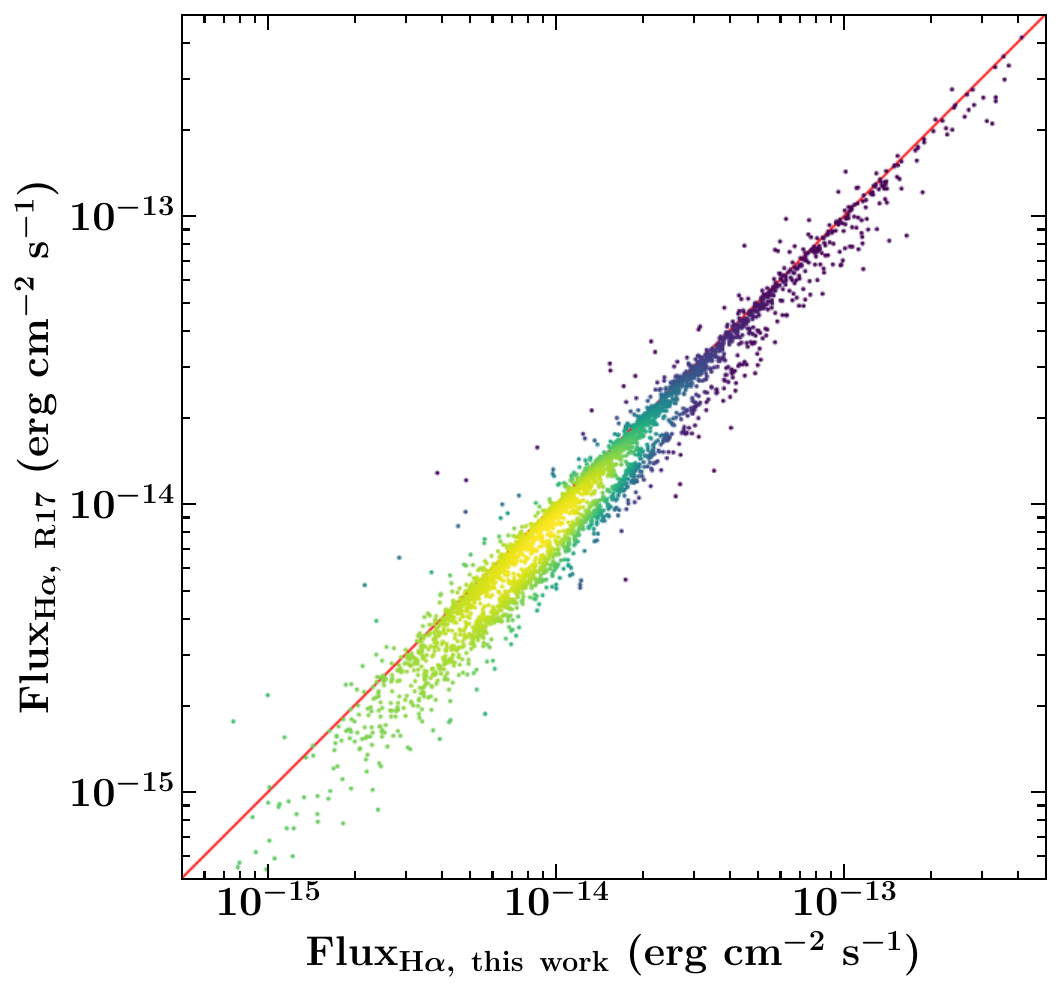}
}
\hbox{
\includegraphics[scale=0.3]{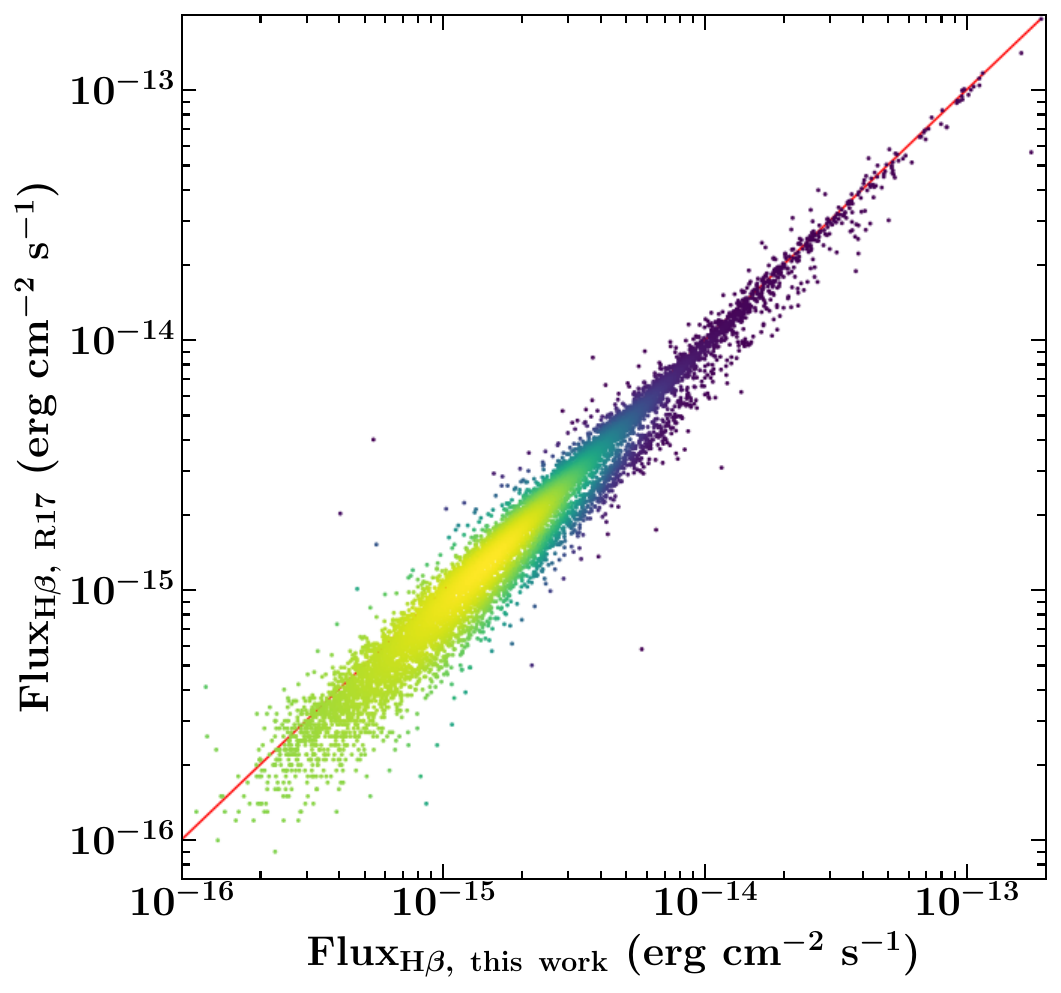}
\includegraphics[scale=0.3]{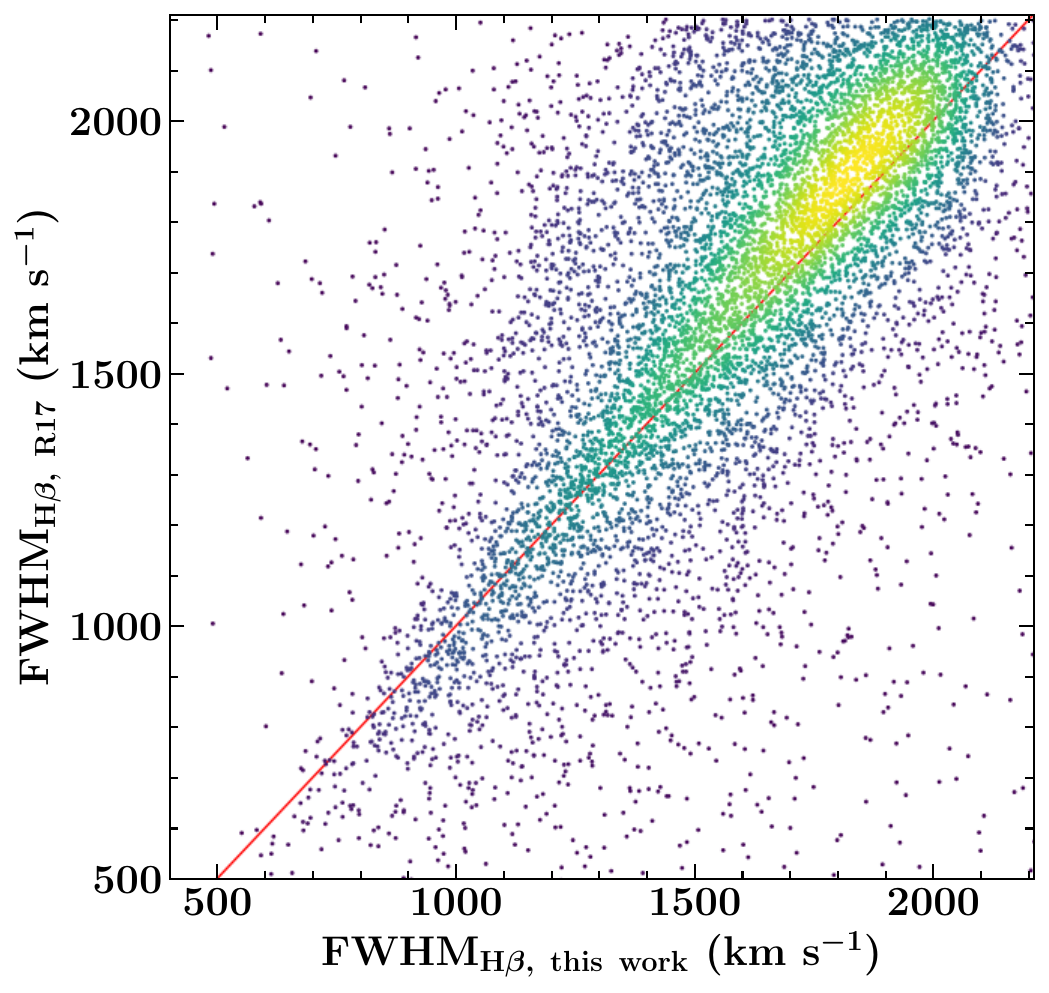}
\includegraphics[scale=0.3]{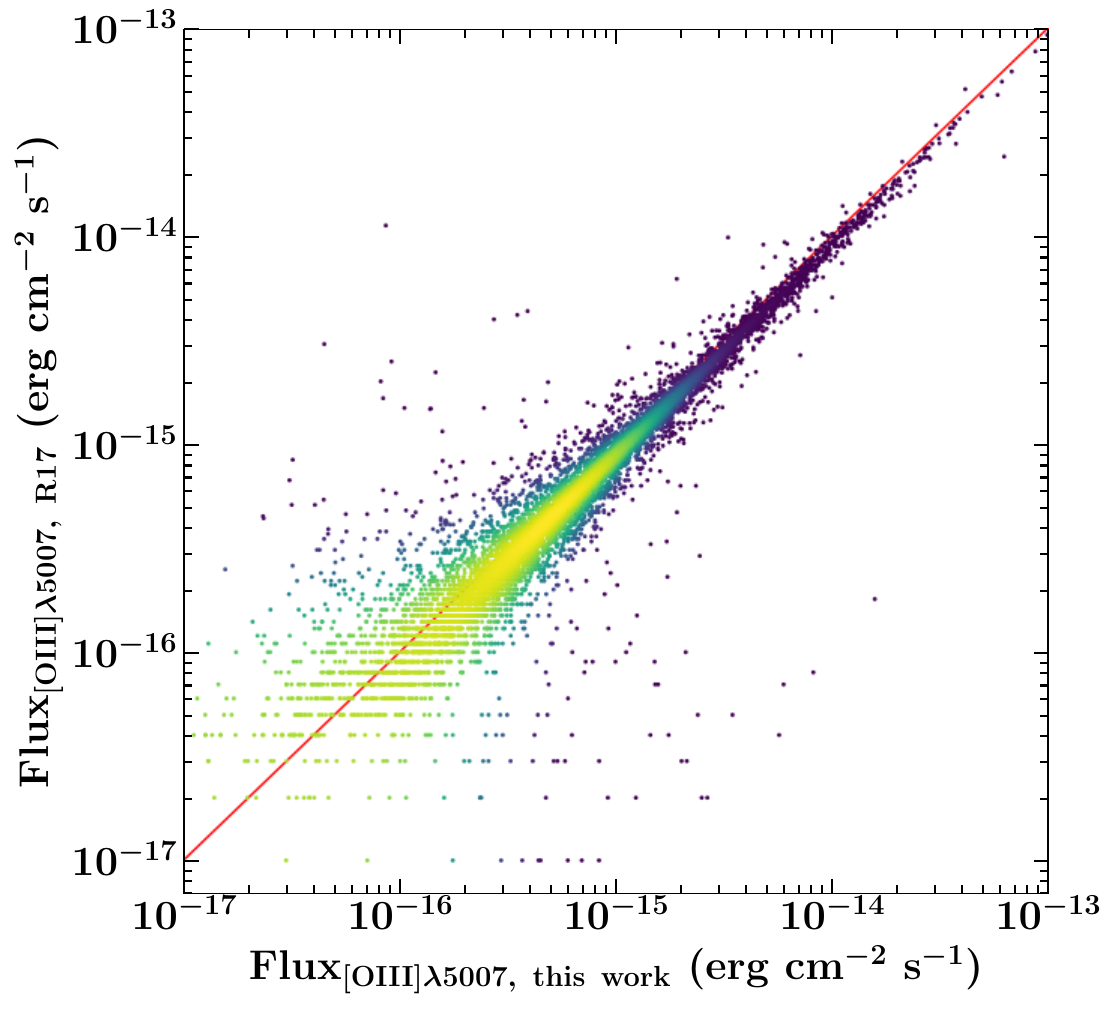}
}
\hbox{
\includegraphics[scale=0.3]{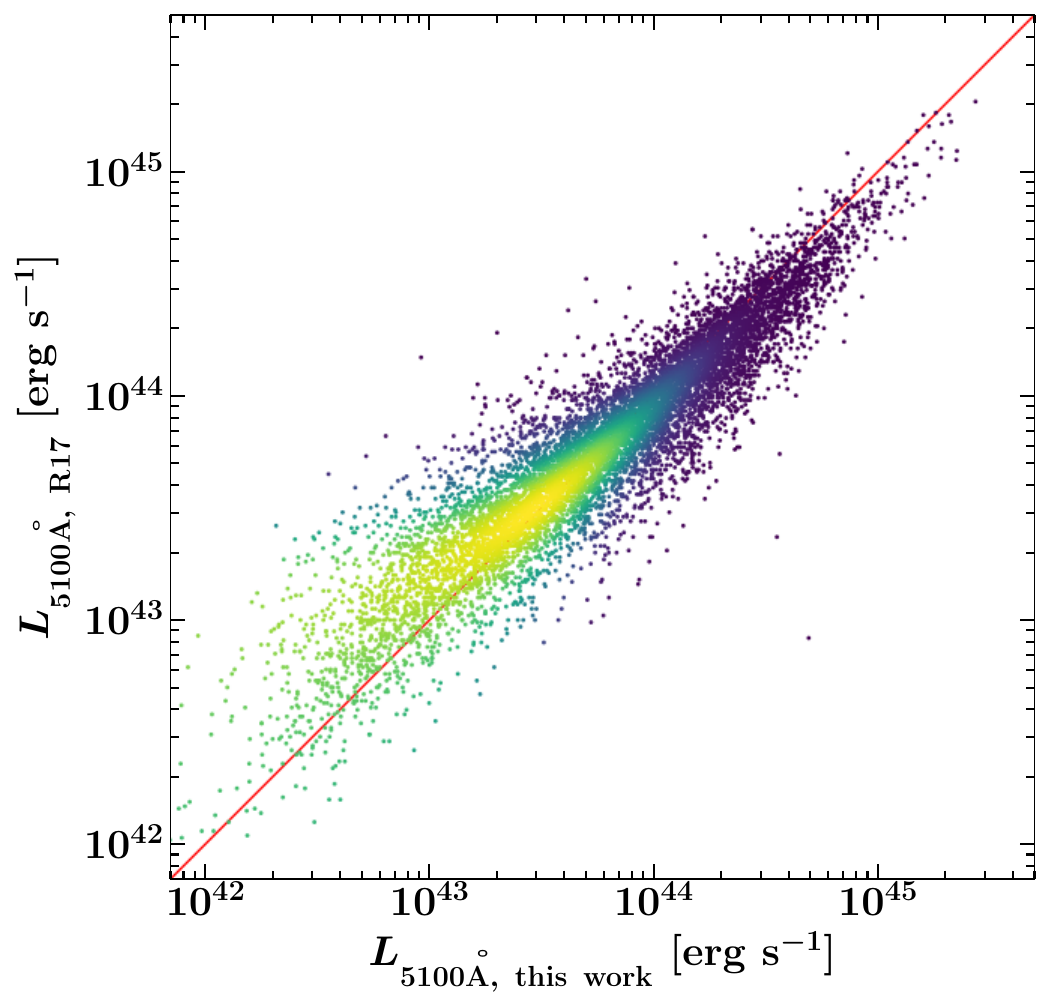}
\includegraphics[scale=0.3]{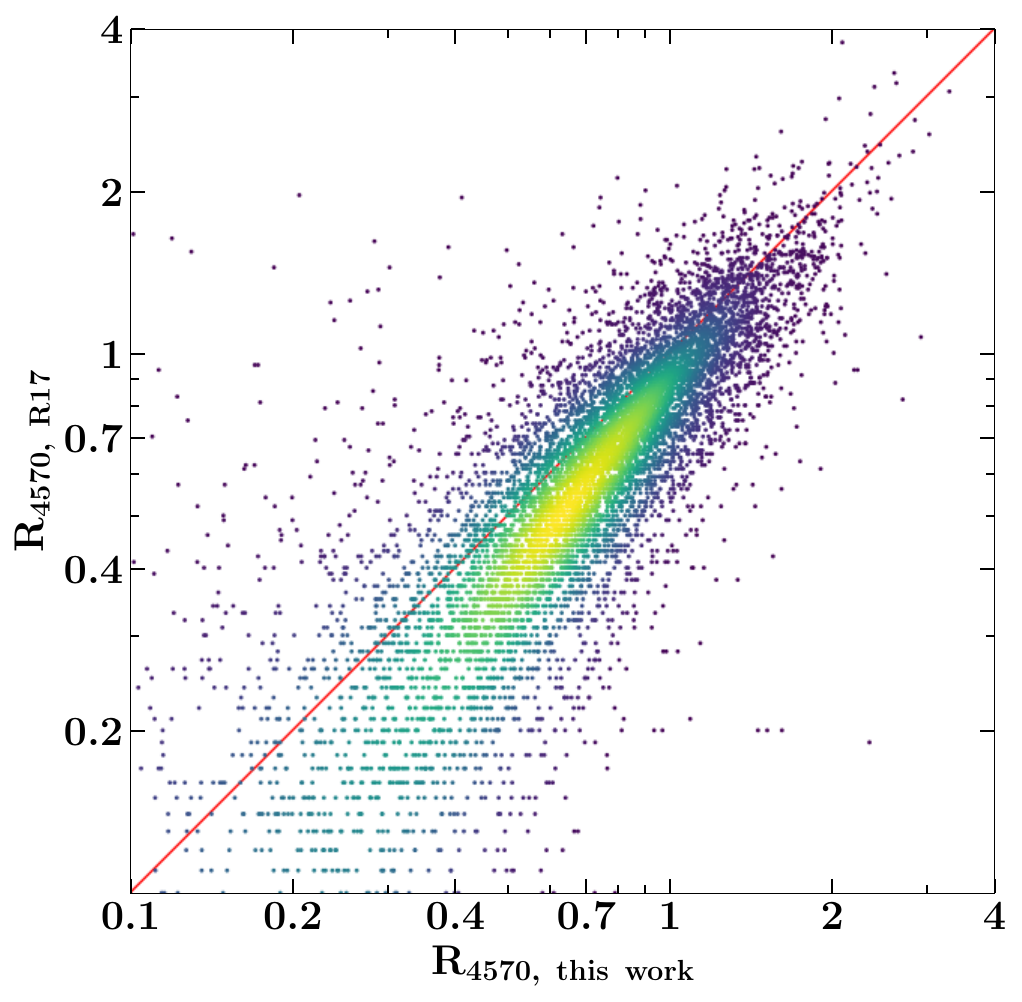}
}
\caption{This plot shows the comparison of various spectral parameters obtained in this work and that published for SDSS-DR12 NLSy1 catalogue \citep[R17,][]{2017ApJS..229...39R}. The colour coding is done based on the number density of sources. The red line refers to the one-to-one correlation.} \label{fig:suvendu}
\end{figure*}

In Figure~\ref{fig:suvendu}, we have shown comparison of some of the parameters derived in the two papers.  The FWHM values of the broad \halpha~emission line were found to be similar to that measured for SDSS-DR12 NLSy1 objects. The logarithmic ratio of the measured quantities is $-0.04\pm0.10$. A small fraction of the sources appear to be systematically shifted with DR12 measurements having higher values (Figure~\ref{fig:suvendu}, top left panel). To investigate the differences, we cross-matched the SDSS-DR12 NLSy1 catalogue with SDSS-DR3 NLSy1 catalogue \citep[][]{2006ApJS..166..128Z} and compared their measurements of the broad \halpha~FWHM values. The top middle panel of Figure~\ref{fig:suvendu} reveals the comparison where the same pattern can be seen.  This observation indicates a possible issue in estimating the broad \halpha~FWHM values for some of the SDSS-DR12 NLSy1 objects. Furthermore, the broad \halpha~flux measurements done by us reasonably matches with that published for SDSS-DR12 NLSy1s (average logarithmic ratio $0.06\pm0.09$, Figure~\ref{fig:suvendu}, top right panel). In this diagram also, a small fraction of objects have systematically lower broad \halpha~flux values as reported in SDSS-DR12 NLSy1 catalogue. These objects turned out to be the same that have systematically offset FWHM of the broad \halpha~line seen in the top left panel of Figure~\ref{fig:suvendu}.  

\begin{figure}
\vbox{
\includegraphics[scale=0.4]{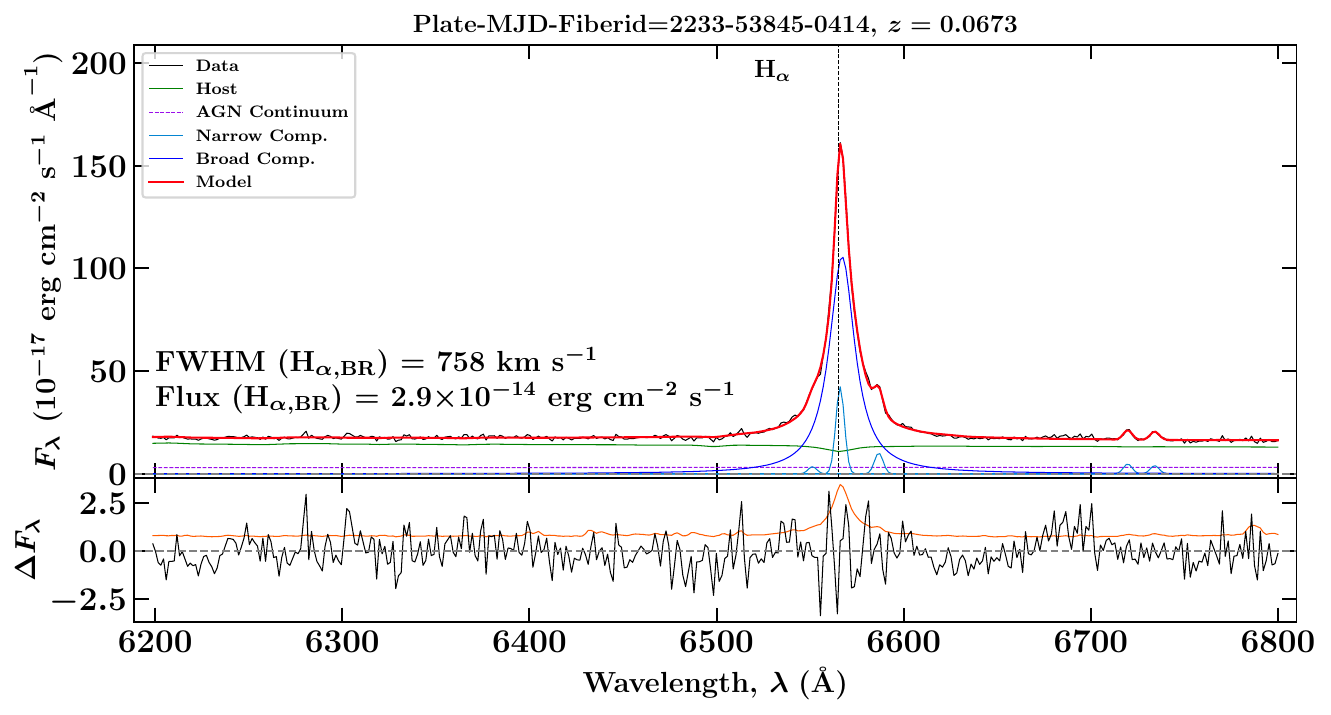}
\includegraphics[scale=0.4]{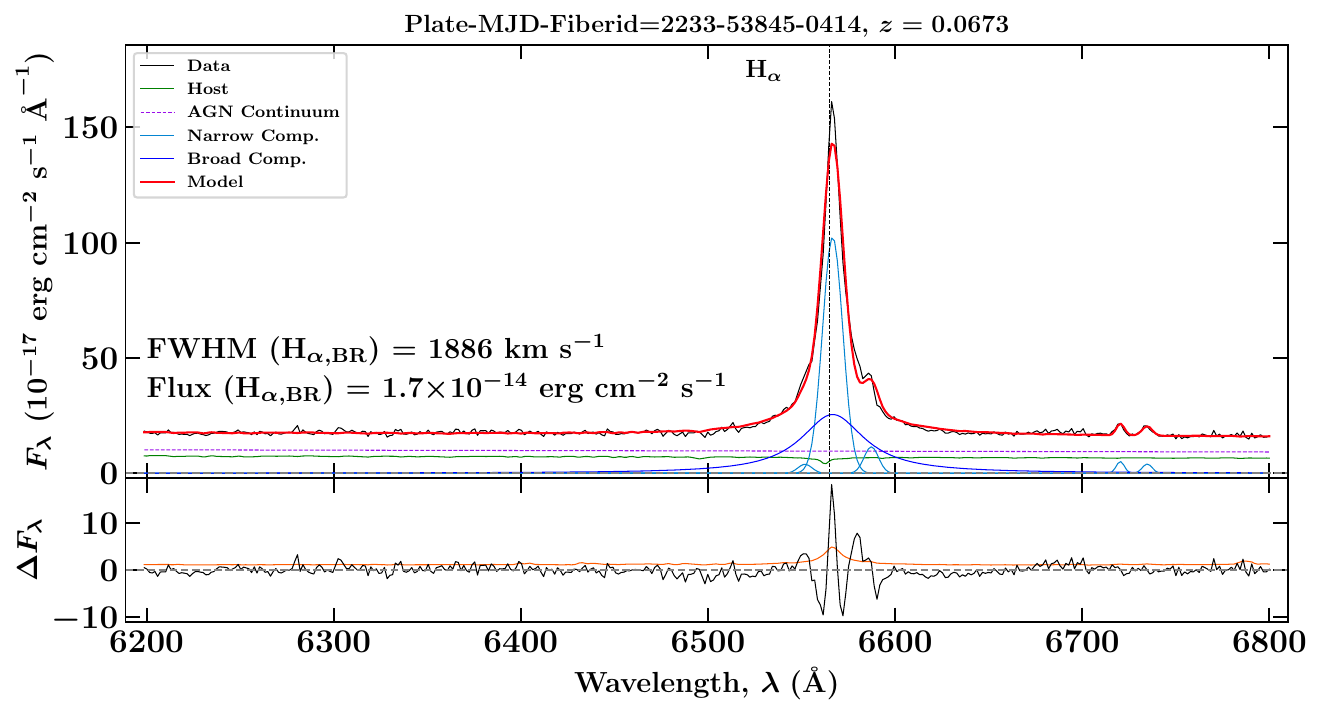}
}
\caption{Results of the spectral fitting done on one of the outliers found in Figure~\ref{fig:suvendu}, top left panel. The top panel shows the fitting result obtained after 5000 iterations of the MCMC fitting, whereas, the bottom panel refers to that obtained for one of the MCMC iterations.  The measured FWHM and flux values for the broad \halpha~component is mentioned. See the text for details.} \label{fig:spec_comp}
\end{figure}

We made an attempt to understand such outliers with larger FWHM and smaller flux values of the broad \halpha~component reported in the SDSS-DR12 NLSy1 catalogue with respect to our measurements. The fitting of the \halpha~line region is complex due to the presence of the \NIIab~doublet overlapping with the broad and narrow \halpha~components. Therefore, one possibility could be due to issues in properly decomposing these emission lines, which can affect the broad \halpha~measurements. In Figure~\ref{fig:spec_comp}, we show the spectral fitting results for one of the outliers. The top panel shows the final result derived after running 5000 iterations of the MCMC fitting.  We obtained the broad \halpha~FWHM of 758 \km~and the flux of $2.9\times10^{-14}$ \ergflux. The bottom panel shows the fitting result for one of the MCMC iterations where the \halpha~line was mainly fitted with the narrow \halpha~component, thereby suppressing the broad \halpha~component. This led to broader FWHM (1856 \km) and lower flux ($1.7\times 10^{-14}$ \ergflux) values for the broad \halpha~component. Indeed, outliers with larger FWHM have lower flux values in the SDSS-DR12 NLSy1 catalogue with respect to our measurements (Figure~\ref{fig:suvendu}, top left and right panels). Therefore, it is possible that the \halpha~line region fitting done by \citet[][]{2017ApJS..229...39R} might have found local minima (with fitting solutions similar to the bottom panel of Figure~\ref{fig:spec_comp}) for such outliers.

We found close matches between broad \hbeta~flux and FWHM measurements with the logarithmic ratio distribution having an average of $0.06\pm0.09$ and $-0.01\pm0.10$, respectively (Figure \ref{fig:suvendu},middle left and central panels). The average logarithmic ratio of the \OIIIb~line flux measured in this work to that in SDSS-DR12 is $0.04\pm0.19$. The large dispersion is possibly due to larger scatter seen in fainter sources where two measurements appears to deviate though the overall trend remains the same (Figure~\ref{fig:suvendu}, middle right panel). We have also compared the AGN continuum luminosity at 5100\AA~and found the results to be similar (logarithmic ratio $-0.01\pm0.21$). Finally, a comparison of the \FeII~strength, i.e., $R_{\rm 4570}$ parameter, reveals our measurements to be slightly higher than that found by \citet[][]{2017ApJS..229...39R} for SDSS-DR12 NLSy1 galaxies. The logarithmic ratio of the two measurements is $0.08\pm0.18$.  Altogether, though the spectral modeling strategies of both works are different, the overall results appears similar and we were able to retrieve $>$80\% of SDSS-DR12 NLSy1 galaxies thereby hinting at the robustness of the fitting procedure.  

\begin{figure*}
\hbox{\hspace{0cm}
\includegraphics[scale=0.3]{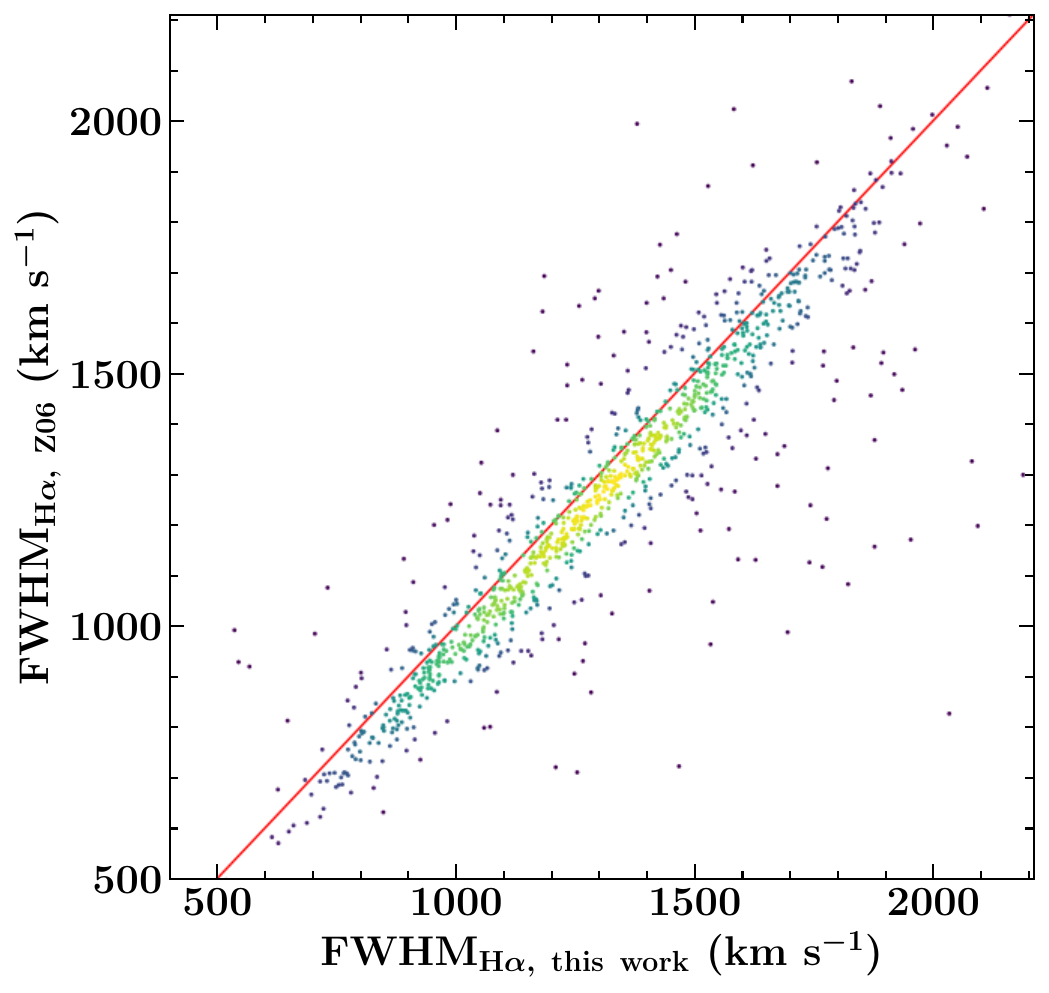}
\includegraphics[scale=0.3]{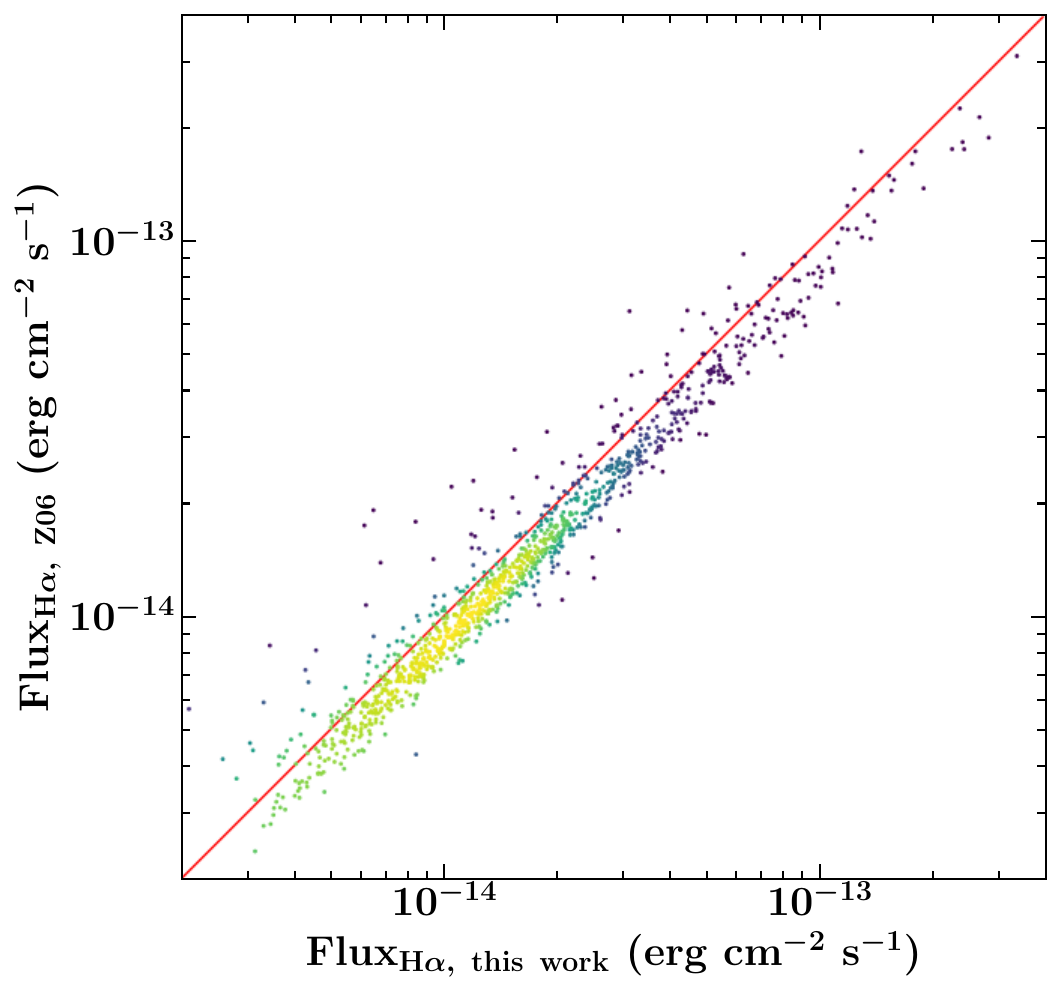}
\includegraphics[scale=0.3]{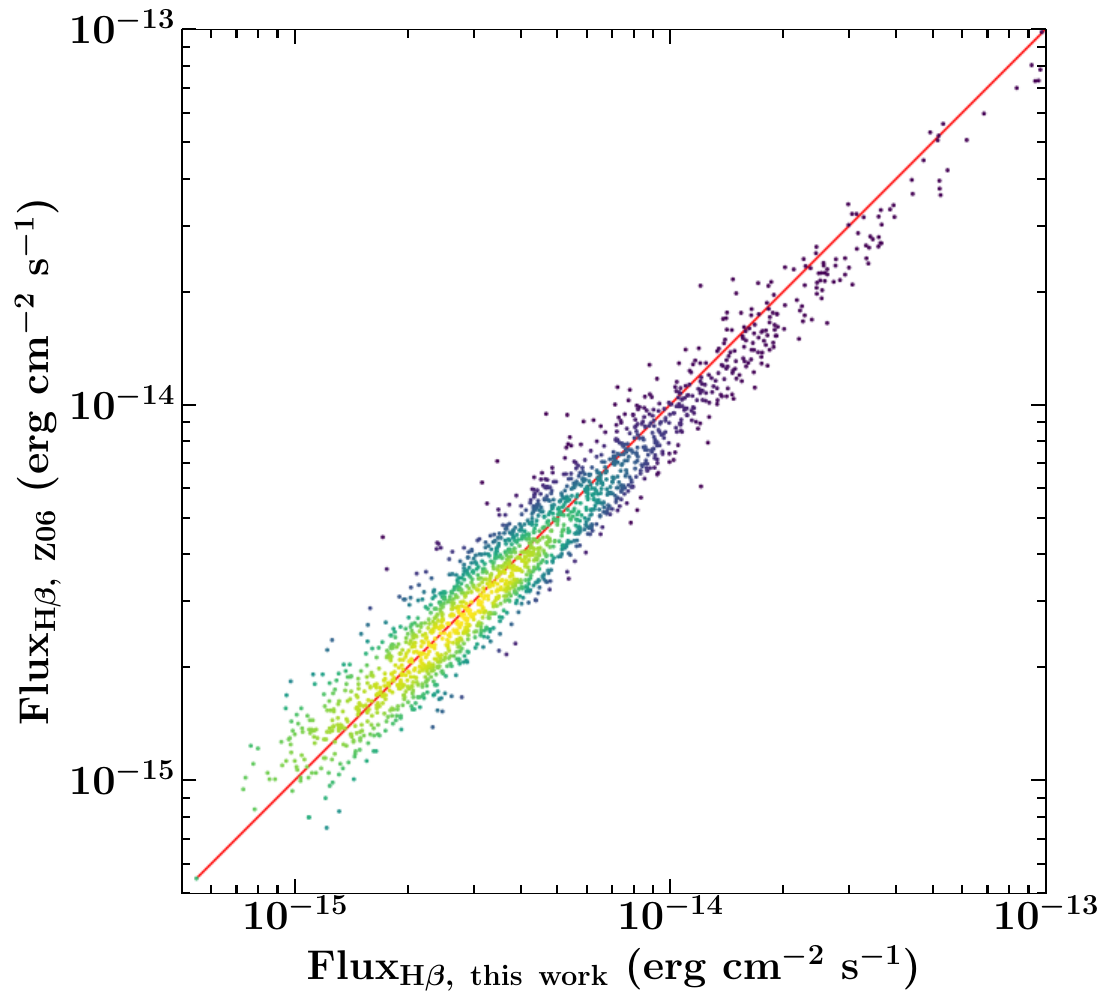}
}
\hbox{
\includegraphics[scale=0.3]{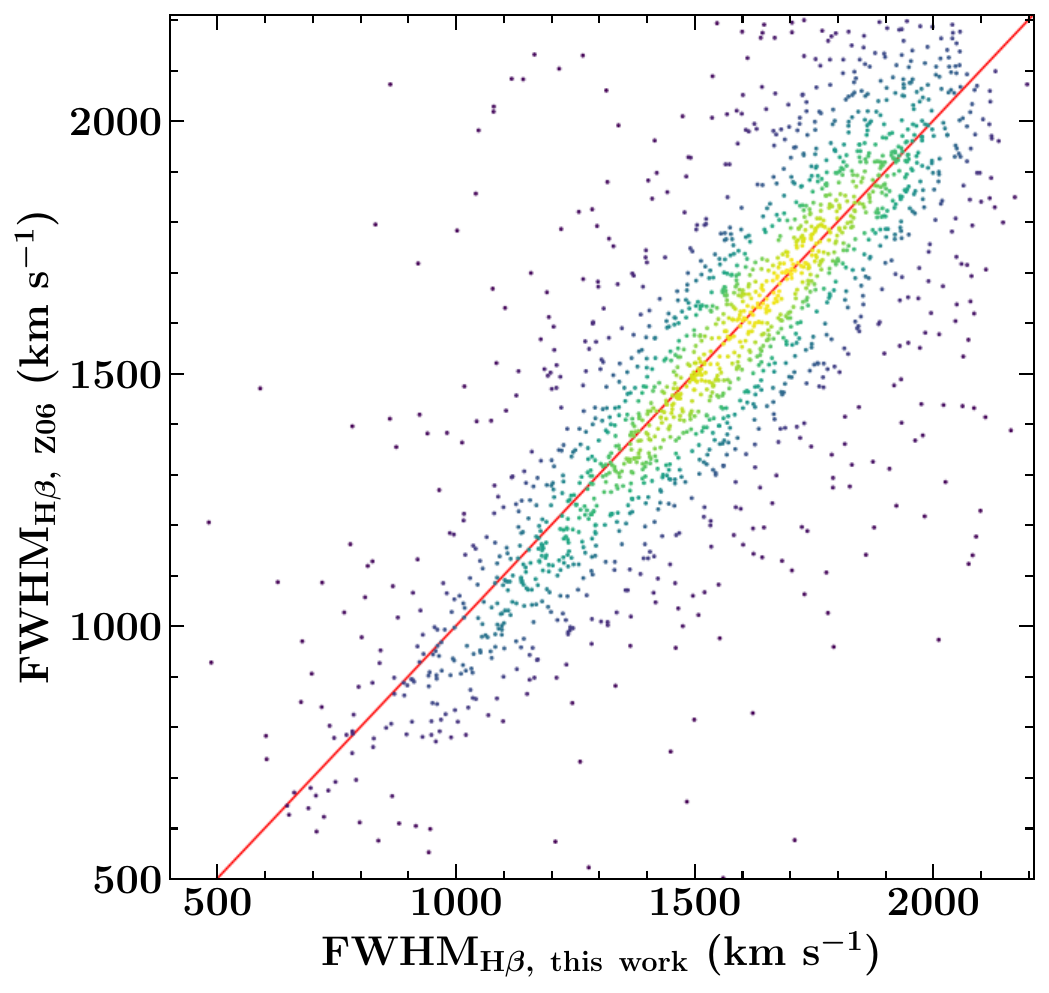}
\includegraphics[scale=0.3]{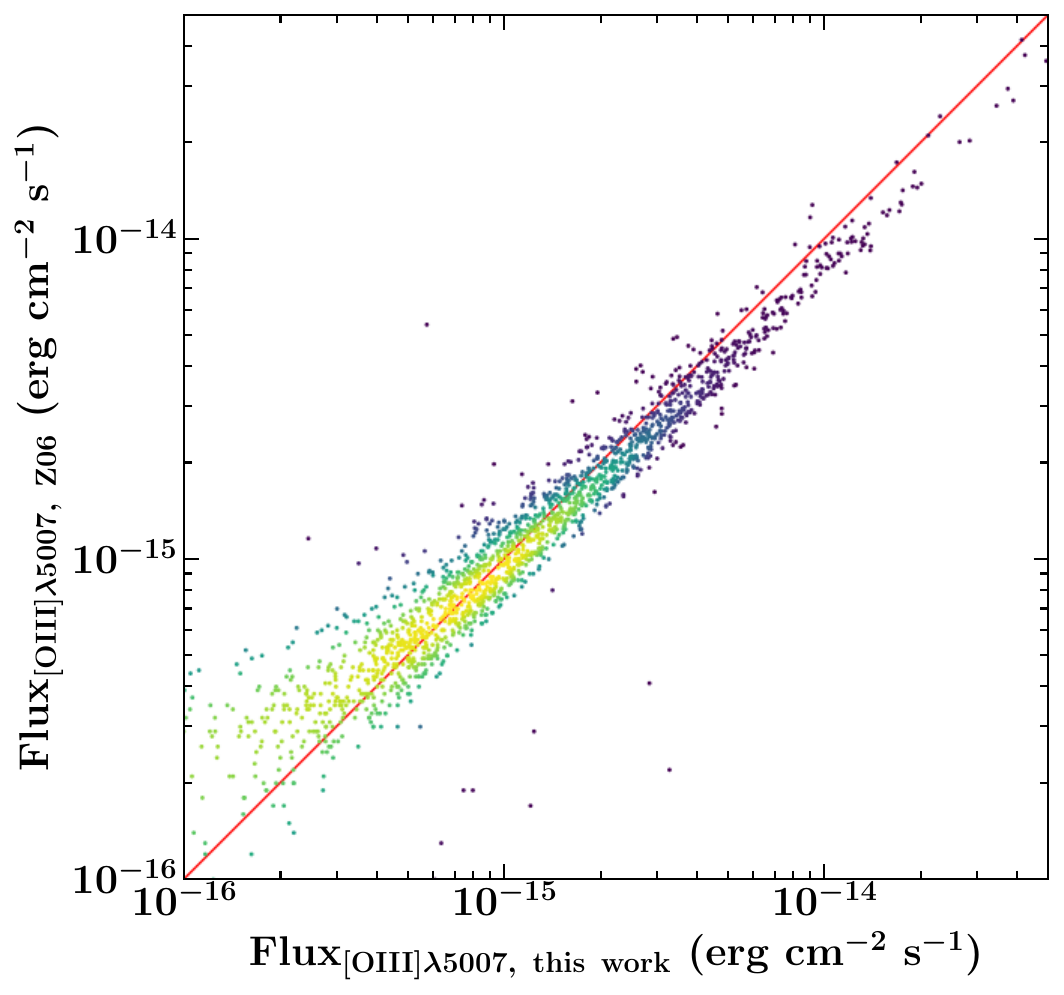}
\includegraphics[scale=0.3]{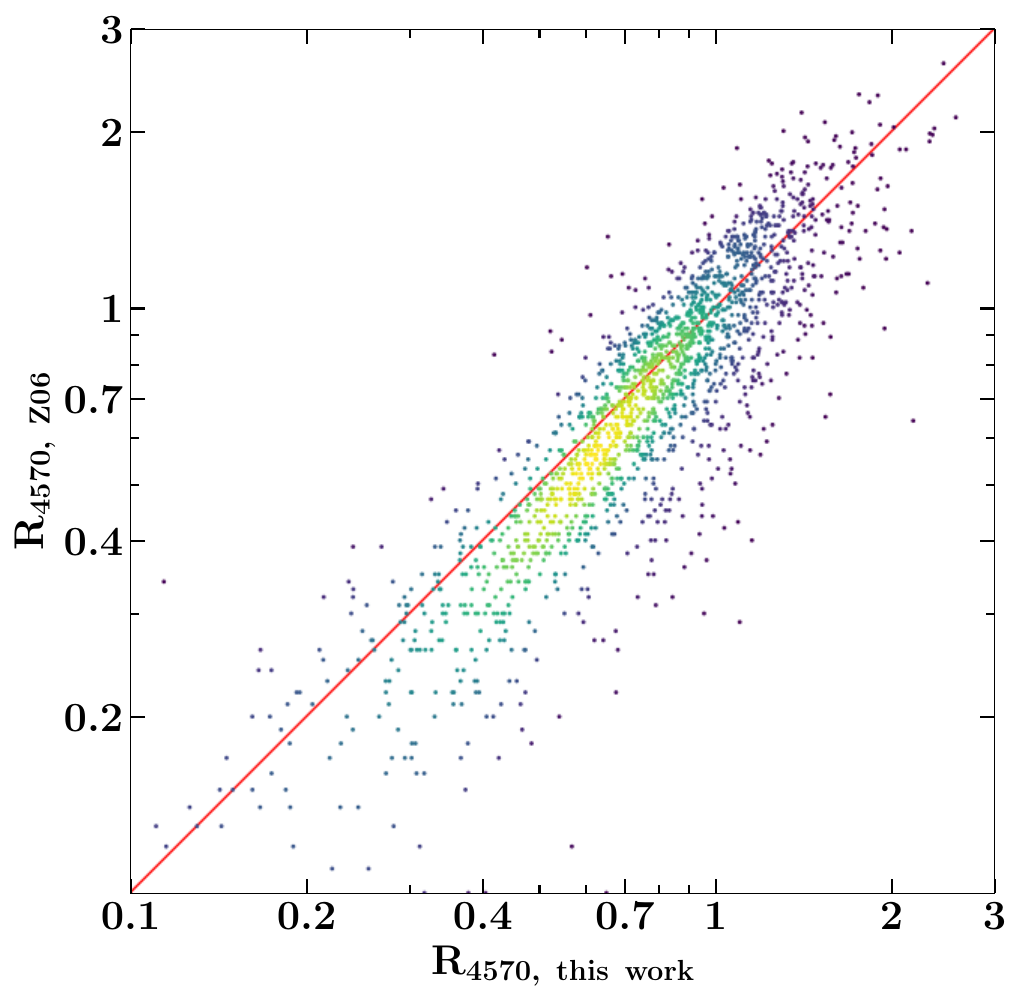}
}
\caption{This plot shows the comparison of various spectral parameters obtained in this work and that published for SDSS-DR3 NLSy1 catalogue \citep[Z06,][]{2006ApJS..166..128Z}.  Other information are same as in Figure~\ref{fig:suvendu}.} \label{fig:zhou}
\end{figure*}

We have also compared our results with the SDSS-DR3 NLSy1 catalogue of 2011 sources \citep[][]{2006ApJS..166..128Z}. There are 23 SDSS-DR3 NLSy1 galaxies which were rejected in our analysis. Among the remaining 1988 objects, 1849, i.e., 93\%, were found common in both the works. Adopting the relaxed threshold of $FWHM_{\rm H\beta}=$2200 \km,  the number of common AGN increases to 1928 i.e., 97\% of the SDSS-DR3 NLSy1 galaxies. We show the comparison of various spectral parameters derived in this work and that obtained by \citet[][]{2006ApJS..166..128Z} in Figure~\ref{fig:zhou}. The logarithmic ratio of the FWHM of the broad \halpha~emission line has an average of $0.01\pm0.05$ and unlike SDSS-DR12 NLSy1s, there is no significant offset of sources (Figure~\ref{fig:zhou}, top left panel). The flux values of the broad \halpha~and \hbeta~emission lines were also similar with the average logarithmic ratio of $0.05\pm0.07$ and $0.01\pm0.08$, respectively (Figure~\ref{fig:zhou}, top middle and right panels). The FWHM of the broad \hbeta~component measured in both the works has an average logarithmic ratio of $0.01\pm0.07$ indicating a close match (Figure~\ref{fig:zhou}, bottom left panel). Similarly, a comparison of the \OIIIb~line flux and \FeII~complex strengths reveal the measured values to be very similar with average logarithmic ratio of $-0.004\pm0.160$ and $0.05\pm0.14$, respectively. All these results highlight the accuracy of the adopted fitting technique and the robustness of the derived spectral parameters.

\begin{figure*}
\vbox{
\includegraphics[scale=0.4]{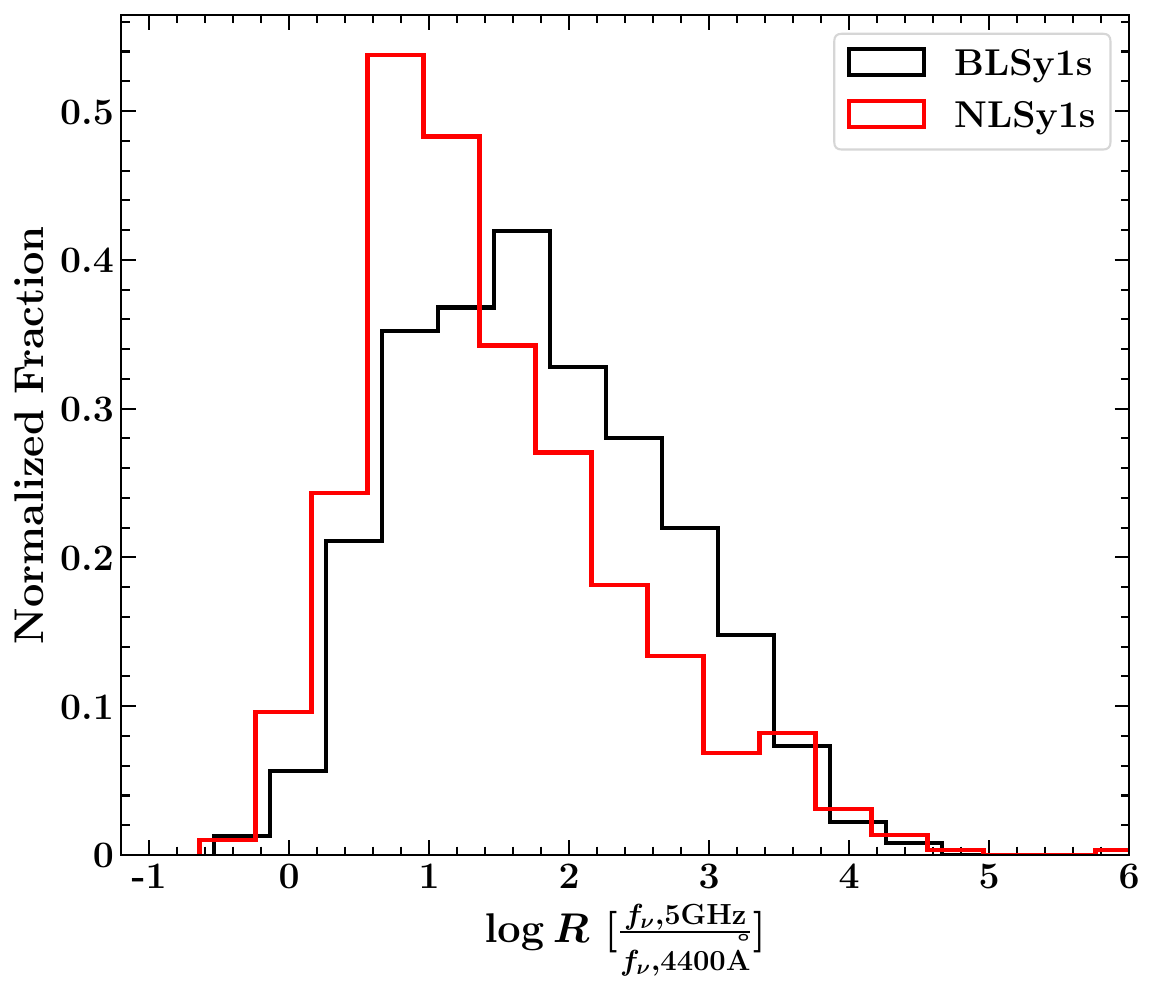}
\includegraphics[scale=0.41]{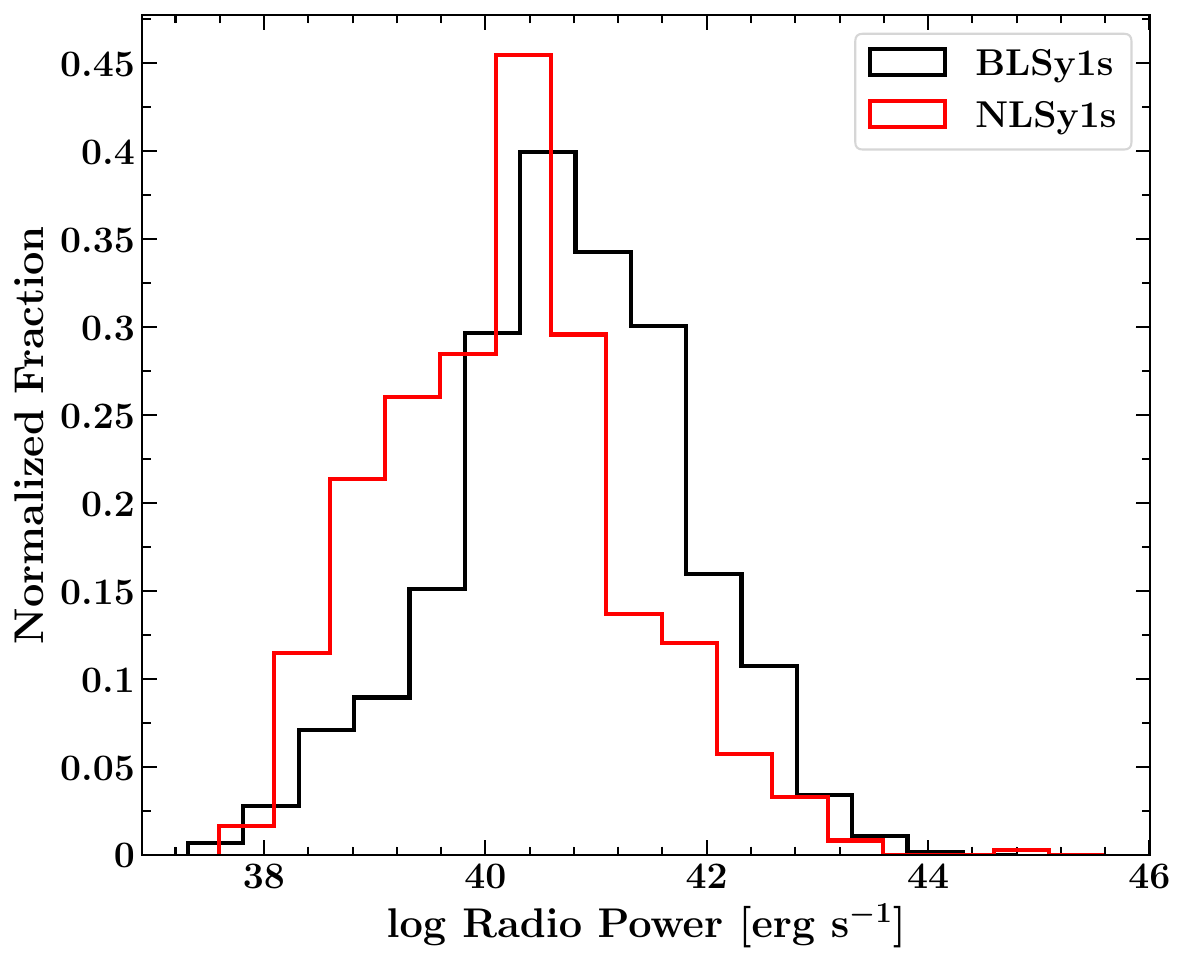}
}
\caption{A comparison of the radio-loudness parameter ($R$, left) and $k$-corrected 1.4 GHz luminosity (right) for NLSy1 and BLSy1 galaxies.} \label{fig:rl_p}
\end{figure*}

\subsection{The BLSy1 Catalogue}
The BLSy1 catalogue was prepared from the parent sample after removing the NLSy1 galaxies and 46 narrow line objects that have the flux ratio of \OIIIb~and \hbeta~emission lines $>$3. The total number of BLSy1 galaxies is 52273. We caution that it may not be appropriate to consider all of them as genuine Seyfert galaxies since luminous broad line quasars are also present in the sample. Indeed, dividing the objects based on their $M_{\rm B}$ values, we found 24401 broad line sources to be quasars with $M_{\rm B}<-23$ and 27761 broad line Seyferts with absolute $B$-band magnitude $>-23$ \citep[][]{1983ApJ...269..352S}.  

\section{Multi-wavelength properties}\label{sec5}
The NLSy1 galaxies exhibit peculiar observational characteristics across the electromagnetic spectrum as outlined in Section~\ref{sec1}. Though a detailed multiwavelength study of these sources is beyond the scope of the current work, we briefly describe interesting observational features by cross-matching our catalogues with several multi-frequency catalogues.

\subsection{Radio observations}
The NLSy1 objects are generally faint radio emitters with only a small fraction ($\sim$5\%) detected in radio surveys \citep[cf.][]{2006AJ....132..531K,2018MNRAS.480.1796S}. We cross-matched our NLSy1 sample with the Faint Images  of the Radio Sky at Twenty-Centimeters \citep[FIRST,][]{1997ApJ...475..479W} with 5 arcsec search radius and found 730 radio emitting NLSy1 galaxies. Interestingly, though our catalogue is more than two times larger than SDSS-DR12 NLSy1 catalogue, the number of radio detected sources has increased only by a factor of $\sim$1.3. The fraction of radio detected NLSy1 galaxies in our sample is $\sim$3\% which is smaller than $\sim$5\% reported for SDSS-DR12 NLSy1 catalogue \citep[][]{2017ApJS..229...39R}.  This can be understood by considering the fact that the enhancement in the number of NLSy1s is largely at higher redshifts (Figure~\ref{fig:z_dis}). If we assume the radio luminosity of the sources to be similar, a higher redshift implies a fainter flux which may remain below the sensitivity of the FIRST survey.

We derived the radio-loudness parameter ($R$) as the ratio of the rest-frame flux densities at 5 GHz and 4400 \AA~\citep[][]{1989AJ.....98.1195K}. The 5 GHz flux density was calculated by extrapolating the 1.4 GHz integrated flux density assuming the spectral index $\alpha=0.5$ ($F_{\nu}\propto\nu^{-\alpha}$). Out of the 730 radio detected NLSy1s, 460 are identified as radio-loud ($R>10$). This fraction (460/730, $\sim$63\%) is smaller than the $\sim$68\% found for SDSS-DR12 NLSy1 objects \citep[][]{2017ApJS..229...39R}.  The difference in the adopted methods to compute the $R$ parameter in both works could explain this observation. Indeed, \citet[][]{2017ApJS..229...39R} calculated it as the ratio of the flux densities at 1.4 GHz and optical $g$-band flux, whereas, we have used the conventional definition \citep[][]{1989AJ.....98.1195K}.

The cross-matching of the BLSy1 catalogue with the FIRST survey led to the identification of 2568 radio detected AGN among which 1975, i.e., $\sim$77\% are found to have $R>10$. On comparing the radio-loudness with that estimated for NLSy1 sample, the overall distributions appear similar (logarithmic dispersion $\sim$0.9, Figure~\ref{fig:rl_p}) though the median average $R$ value for BLSy1s ($\sim$51) is higher than that estimated for NLSy1 galaxies ($\sim$17). Similarly, the distributions of the 1.4 GHz radio power for both populations also have a similar dispersion ($\sim$1.1, on logarithmic scale) though BLSy1 galaxies are more luminous (median average $\sim5.7\times10^{40}$ \lum) compared to NLSy1 sources ($\sim1.6\times10^{40}$ \lum). The distributions are plotted in the right panel of Figure~\ref{fig:rl_p}.

\begin{figure}
\vbox{
\includegraphics[width=\linewidth]{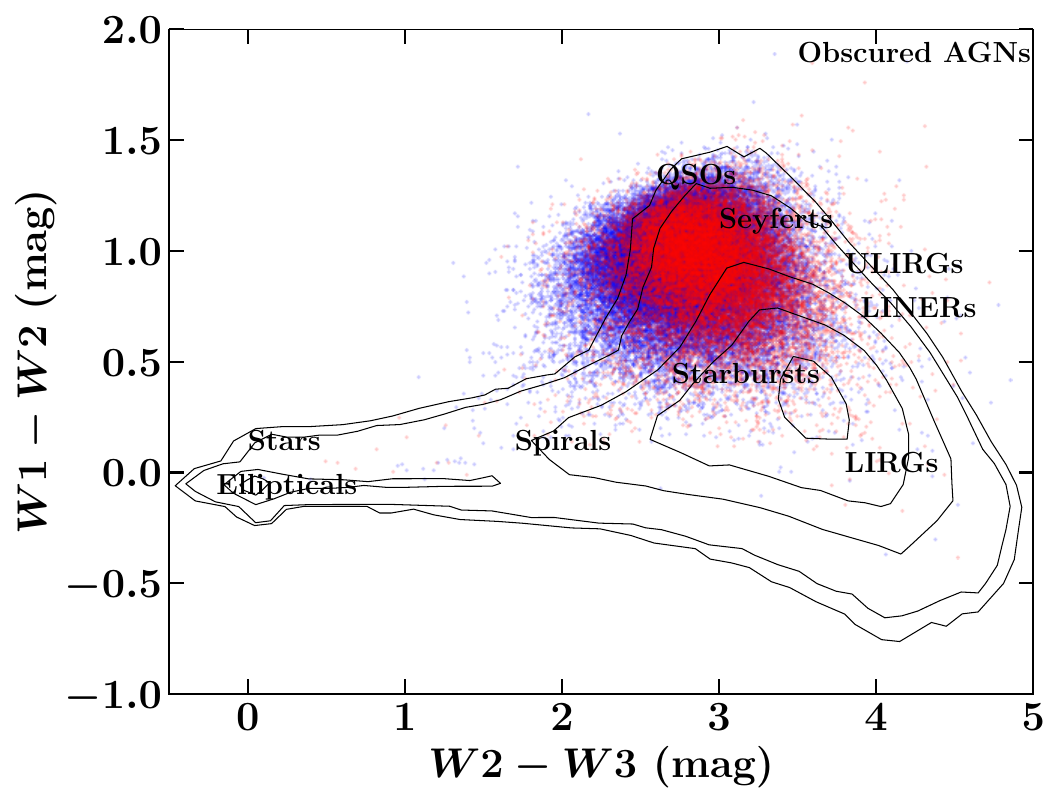}
}
\caption{WISE colour-colour diagram for objects studied in this work. The plotted isodensity contours refer to WISE thermal sources and locations of various source classes are also highlighted. The acronyms QSOs, ULIRGs, LIRGs, and LINERs refer to quasars, ultraluminous infrared galaxies,  luminous infrared galaxies, and low-ionization nuclear emission region galaxies, respectively. The contour data are adopted from \citet[][]{2011ApJ...740L..48M}.} \label{fig:wise}
\end{figure}

\begin{figure}
\vbox{
\includegraphics[width=\linewidth]{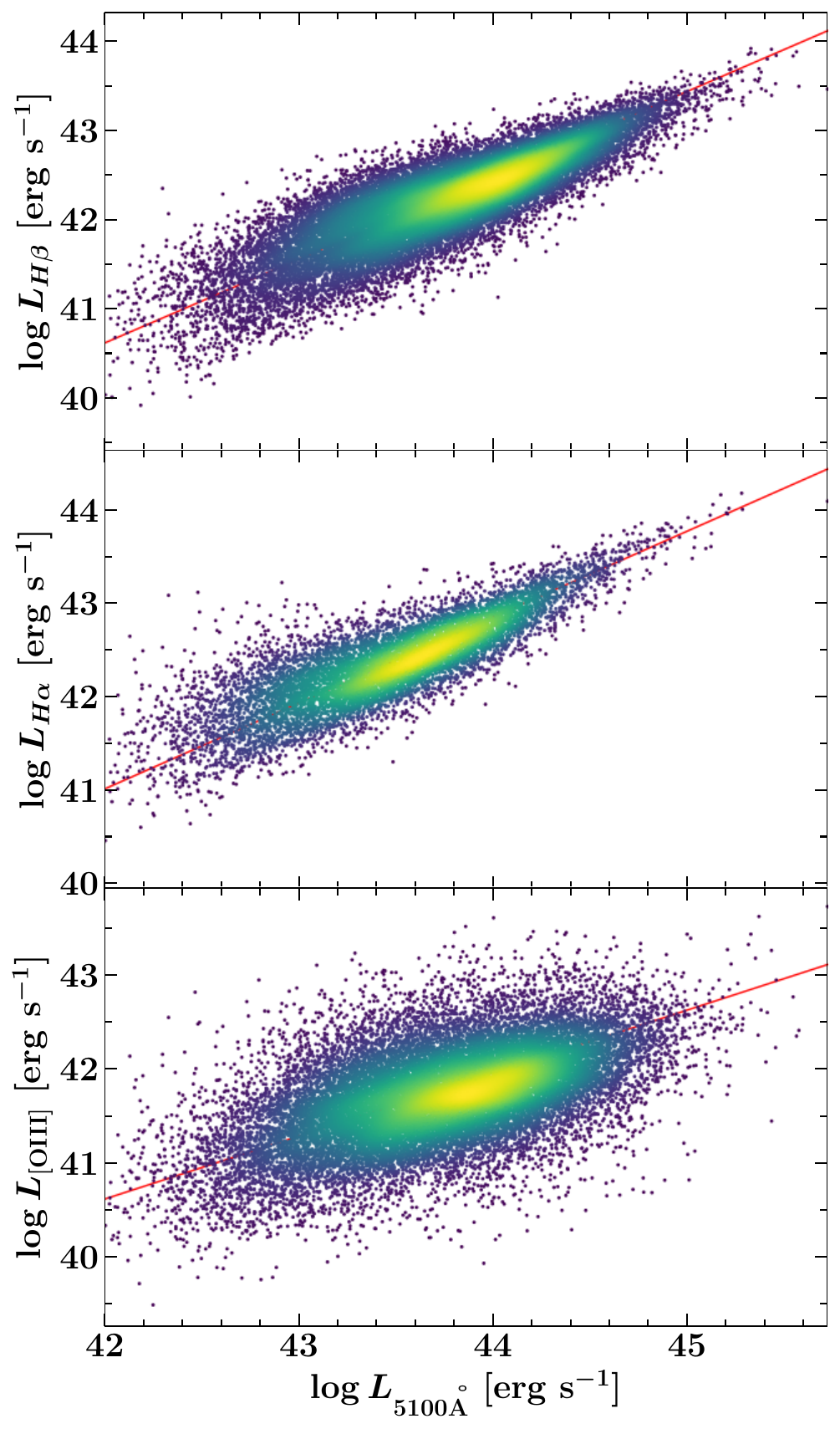}
}
\caption{A comparison of the luminosity of the emission lines and 5100\AA~continuum luminosity . The colour coding is done based on the number density of sources. The red line refers to the best-fitted correlation. See the text for details.} \label{fig:L_L}
\end{figure}

\begin{figure*}
\vbox{
\includegraphics[scale=0.25]{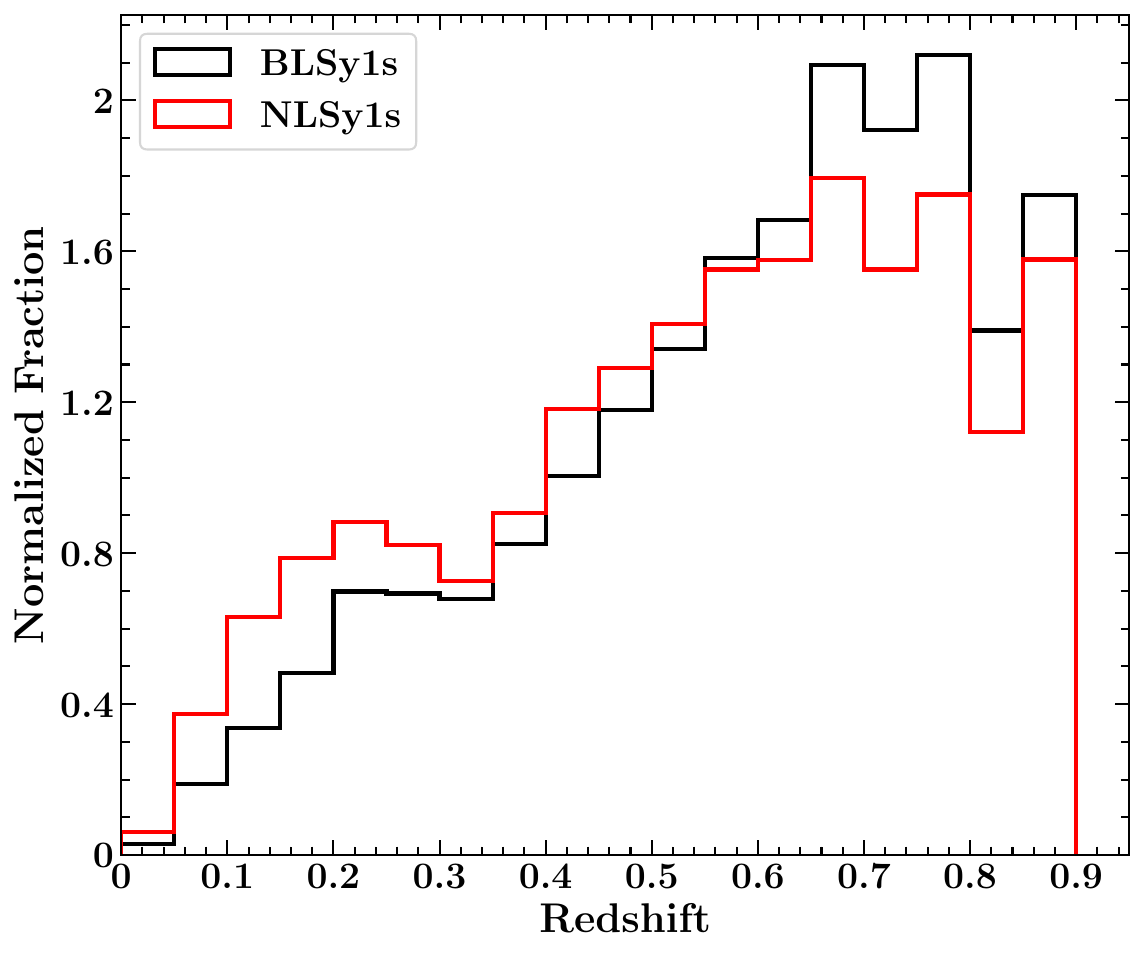}
\includegraphics[scale=0.25]{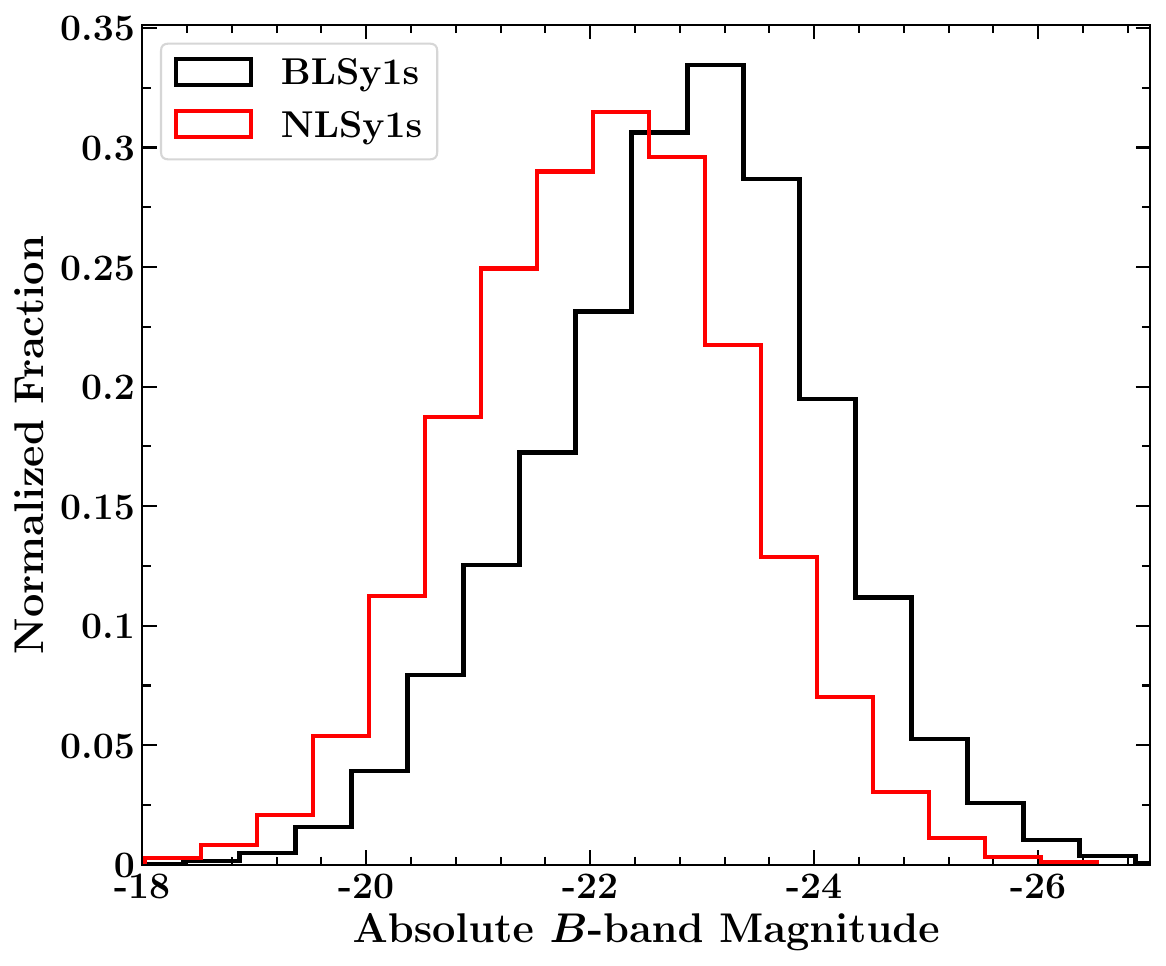}
\includegraphics[scale=0.25]{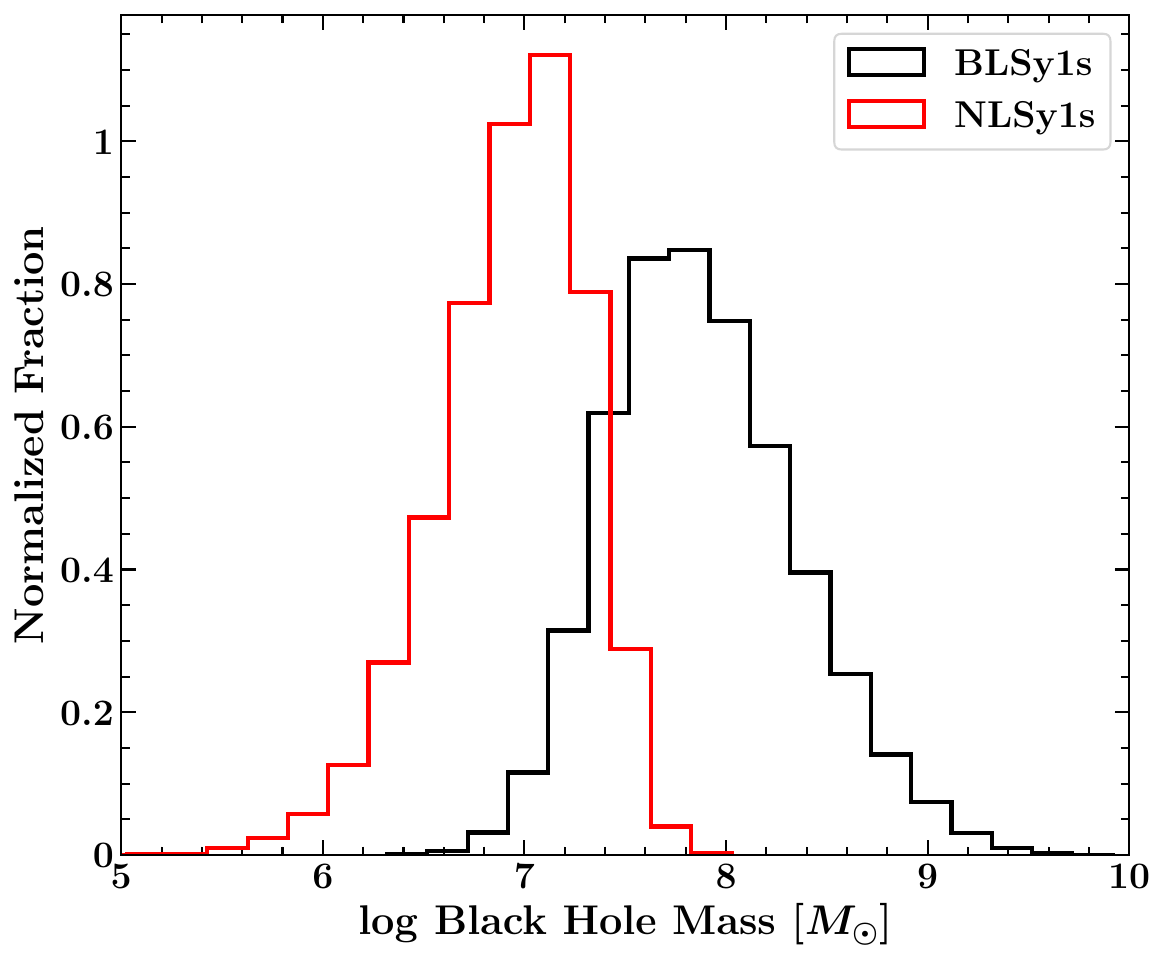}
}
\vbox{
\includegraphics[scale=0.25]{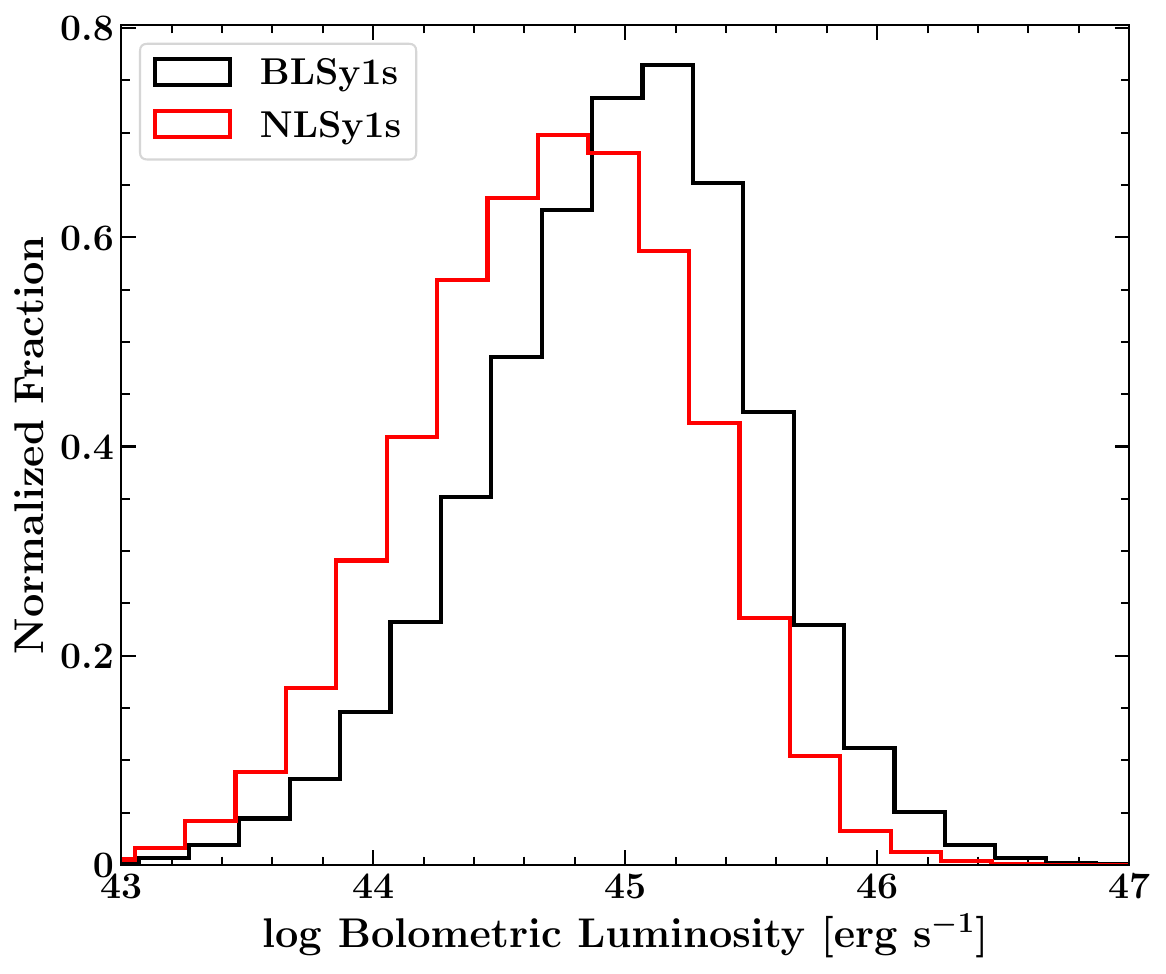}
\includegraphics[scale=0.25]{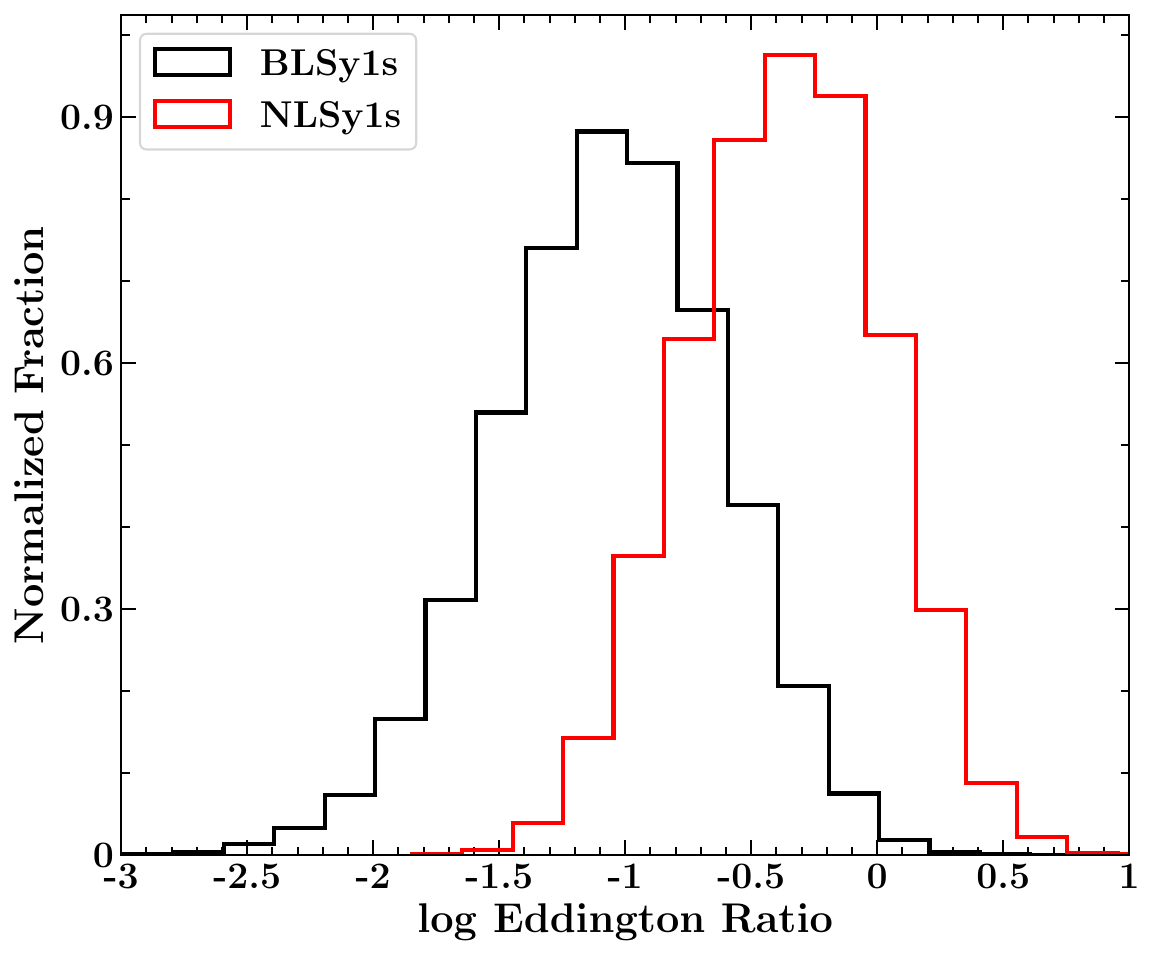}
\includegraphics[scale=0.25]{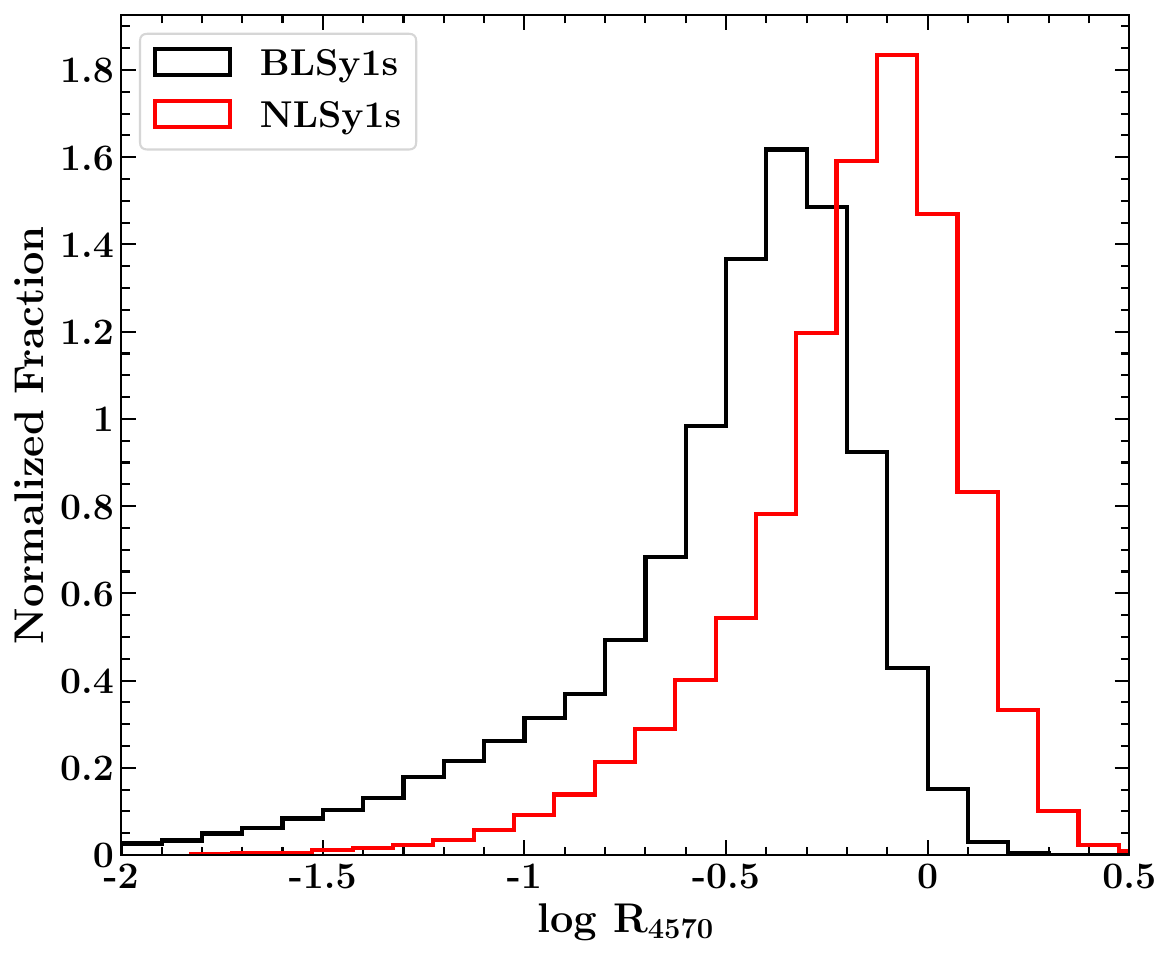}
}
\caption{Comparisons of various optical spectroscopic parameters, measured/derived, for NLSy1 and BLSy1 sources. See the text for details.} \label{fig:optical}
\end{figure*}

\subsection{Infrared observations}
The mid-infrared counterparts of the NLSy1/BLSy1 sources were identified by considering the Wide-field Infrared Survey Explorer \citep[WISE,][]{2010AJ....140.1868W} catalogue and applying a search radius of 5 arcsec. Out of 22656 NLSy1 objects, 21825 ($\sim$96\%), 21807 ($\sim$96\%), and 18118 ($\sim$80\%) are detected in $W1$, $W2$, and $W3$ filters, respectively, at $>$3$\sigma$ confidence level. Considering 52273 BLSy1 galaxies,  51229 ($\sim$98\%), 51231 ($\sim$98\%), and 45759 ($\sim$88\%) are detected in the same bands, respectively.

We show the NLSy1 and BLSy1 objects detected in all three $W1$, $W2$, and $W3$ bands in the WISE colour-colour diagram in Figure~\ref{fig:wise}. Both BLSy1 and NLSy1s occupy overlapping regions which is found to be dominated by luminous quasars and Seyfert galaxies \citep[cf.][]{2013AJ....145...55Y,2018ApJS..234...23A}. A small fraction of sources also spread into the region of starburst galaxies.

\subsection{Optical observations}
A strong correlation between continuum luminosity and emission line luminosities has been reported both for quasar and NLSy1 populations covering a wide range of redshift and bolometric luminosities \citep[][]{2005ApJ...630..122G,2006ApJS..166..128Z,2015ApJ...806..109J,2017ApJS..229...39R}. We show the variations of the \halpha, \hbeta, and \OIIIb~emission line luminosities as a function of the 5100\AA~continuum luminosity for NLSy1 sources in Figure~\ref{fig:L_L}.  The following relations were found by applying a linear least-square fit on the data:

\begin{equation}
\log(L_{\mathrm{H\beta}})=(1.13 \pm 0.12)+ (0.938 \pm 0.002) \log L_{5100}.
\end{equation} 
\begin{equation}
\log(L_{\mathrm{H\alpha}})= (2.47 \pm 0.17) + (0.917 \pm 0.004) \log L_{5100}.
\end{equation} 
\begin{equation}
\log(L_{\mathrm{[O {\small III}]}})= (12.44 \pm 0.22) + (0.675 \pm 0.005) \log L_{5100}.
\end{equation} 

All correlations were found $>$5$\sigma$ significant as also reported for SDSS-DR12 NLSy1 catalogue \citep[][]{2017ApJS..229...39R}. These correlations between the continuum and emission line luminosities enable us to adopt the latter while estimating virial black hole masses for the cases where measuring the former is tedious, e.g., in blazars, galaxy-dominated low-luminosity sources, and Type 2 AGN \citep[][]{2003AJ....126.2125Z,2005ApJ...630..122G}.

We show some of the parameters and physical properties of NLSy1 and BLSy1 sources estimated from the optical spectroscopic analysis in Figure~\ref{fig:optical}. The redshift distribution suggests an increase in number for both populations with redshift. At lower redshifts ($z<0.5$), NLSy1s are more common whereas BLSy1 galaxies dominate at higher redshifts. Comparing the absolute $B$-band magnitudes, NLSy1s tend to appear a bit fainter ($\langle M_{\rm B}\rangle=-22.1$) with respect to BLSy1 objects ($\langle M_{\rm B}\rangle=-22.9$) though their dispersions are similar, $\sim$1.2 magnitudes. The black hole masses of NLSy1s ($\langle \log M_{\rm SE}\rangle=6.98\pm0.37$, \msun) were also found to be lower than that of BLSy1 galaxies ($\langle \log M_{\rm SE}\rangle=7.85\pm0.47$, \msun). This is likely due to the adopted FWHM of the broad \hbeta~line threshold which is used to compute $M_{\rm SE}$. Furthermore, since the differences in the bolometric luminosity distributions for both types of sources are not significant ($\langle \log L_{\rm bol}\rangle=45.01,~44.74$ \lum, for BLSy1 and NLSy1, respectively), a lower $M_{\rm SE}$ also implies a higher Eddington ratio or $R_{\rm Edd}$ for NLSy1 galaxies (Figure~\ref{fig:optical}, bottom middle panel). These results are consistent with those reported for SDSS-DR12 NLSy1 catalogue \citep[][]{2017ApJS..229...39R}.

NLSy1 sources typically exhibit strong \FeII~emission which is quantified using $R_{\rm 4570}$ \citep[ e.g.,][]{2001AJ....122..549V}. We found the median $R_{\rm 4570}$ parameter value to be 0.71 and 0.38 and 1$\sigma$ dispersion to be 0.41 and 0.24 for NLSy1 and BLSy1 galaxies, respectively. These quantities are similar to those reported by \citet[][]{2017ApJS..229...39R} and follow the same trend observed in earlier studies \citep[cf.][]{1984MNRAS.207..263B,2001AJ....122..549V,2006ApJS..166..128Z}.

\subsection{X-ray observations}
\begin{figure}
\vbox{
\includegraphics[width=\linewidth]{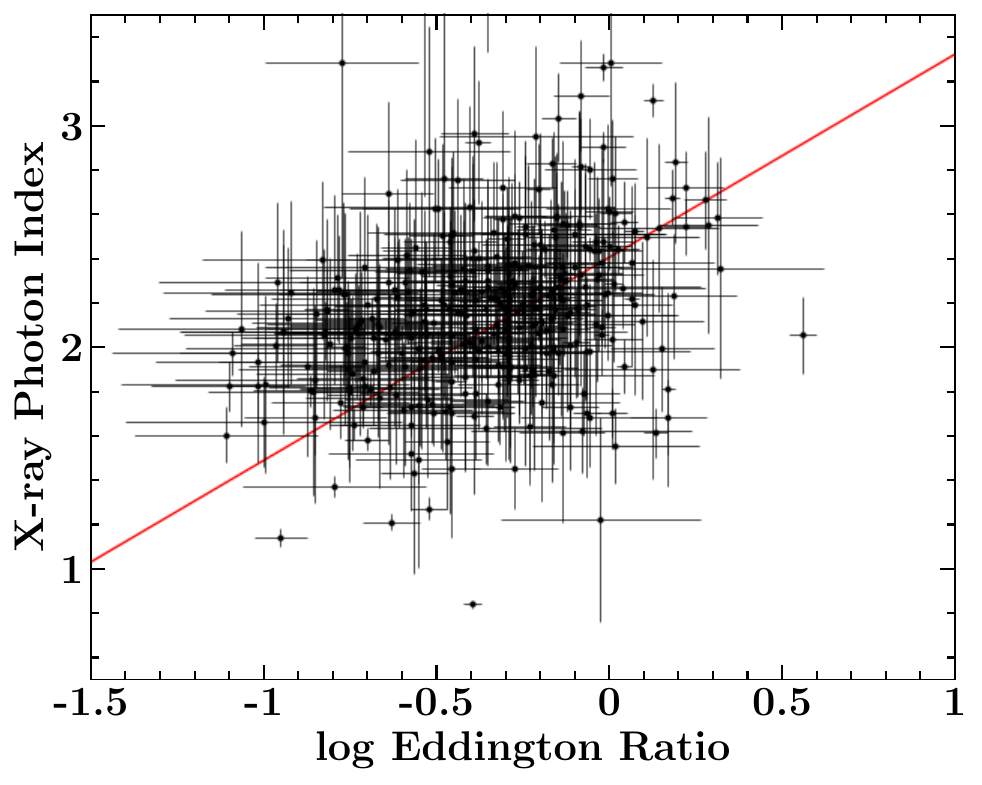}
}
\caption{The variation of the X-ray photon index as a function of the Eddington ratio for NLSy1 galaxies detected with \chandra~and eROSITA satellites. The red line corresponds to the best-fitted correlation.} \label{fig:xray}
\end{figure}

We cross-matched our NLSy1 catalogue with the \chandra~source catalogue \citep[CSC v2.0,][]{2010ApJS..189...37E}, the eROSITA Final Equitorial Depth Survey AGN catalogue \citep[eFEDS,][]{2022A&A...661A...5L}, \xmm~serenditipious source catalogue \citep[4XMM-DR12,][]{2020A&A...641A.136W},  and live \swift-X-ray Telescope Point Source Catalogue \citep[LSXPS,][]{2023MNRAS.518..174E}, in sequential order to identify X-ray emitting sources. Using a search radius of 5 arcsec, we found 567, 195, 979, and 819 matches, respectively. Some of these catalogues, e.g., eFEDS, provide spectral index measurements which we have used to determine its possible correlation with the accretion rate in Eddington units \citep[cf.][]{2009ApJ...700L...6R}. No assumptions were made while considering spectral indices even though these X-ray surveys cover overlapping yet different energy bands.  In Figure~\ref{fig:xray}, we show the variation of the photon index with Eddington ratio for sources detected with \chandra~and eROSITA satellites. A positive correlation is evident with the null hypothesis of no correlation is rejected at $>$5$\sigma$ confidence level. The best-fitted line slope ($m=0.91\pm0.02$) is steeper than that reported in other studies \citep[][]{2009ApJ...700L...6R,2022A&A...657A..57L} which can be understood since no detailed X-ray data reduction was performed by us. Indeed, the X-ray spectra of NLSy1 sources exhibit several interesting features, e.g., soft excess below 2 keV \citep[cf.][]{2009Natur.459..540F}, which should be properly taken into account while estimating spectral parameters.

\subsection{Gamma-ray observations}
The detection of variable \gm-ray emission from some of the radio-loud NLSy1 galaxies have provided strong evidence that even low black hole mass systems can launch powerful relativistic jets \citep[e.g.,][]{2009ApJ...707L.142A}. To identify \gm-ray emitting NLSy1 galaxies in our sample, we considered the fourth data release of the fourth catalogue of \gm-ray sources detected by the \fermi-Large Area Telescope \citep[4FGL-DR4,][]{2022ApJS..263...24A,2023arXiv230712546B}. Adopting the counterpart positions given in this catalogue and using a search radius of 5 arcsec, we found 21 \gm-ray detected NLSy1 galaxies. The list is provided in Table~\ref{tab:lat}. All sources previously known in the \gm-ray band except one are present in the list \citep[][]{2019JApA...40...39P}. The only missing object is the flat spectrum radio quasar GB6 J0937+5008 ($z=0.28$) for which our spectral fitting analysis resulted in the broad \hbeta~line FWHM of $2350.2\pm231.6$ \km~thus classifying it as a broad line AGN.  Among the newly identified NLSy1s, 9 sources are found to be \gm-ray emitters. 

Among the broad line sources, 45 are identified as \gm-ray emitters by cross-matching with the 4FGL-DR4 catalogue (Table~\ref{tab:lat}). Interestingly, there are 10 objects that have absolute $B$-band magnitude $M_{\rm B}>-23$, i.e., they are genuine Seyfert galaxies. This observation strengthens the idea that low-luminosity AGN can also host relativistic jets. 

\section{Summary}\label{sec6}
In this work, we have carried out a detailed analysis of $>$2 million SDSS spectra using the publicly available software \bd~to identify NLSy1 galaxies in the latest SDSS-DR17. By reproducing the Balmer emission lines with the Lorentzian profile and also adopting the multi-component continuum fitting which includes contributions from the optical \FeII~emission, host galaxy, and nuclear power-law AGN radiation, we identified 22656 NLSy1 galaxies with broad \hbeta~emission line FWHM $<$2000 \km~within uncertainties. Furthermore,  $>$75\% of these objects are found to be low-luminosity AGN with $M_{\rm B}>-23$ \citep[][]{1983ApJ...269..352S}.  This exercise also led to a new catalogue of 52273 BLSy1 galaxies. Comparing with previous works \citep[][]{2006ApJS..166..128Z,2017ApJS..229...39R}, we found $>$80\% of their NLSy1s to be present in our catalogue {and the corresponding optical spectroscopic parameters were also found to be comparable,} thereby supporting the robustness of our analysis procedure and results. Based on the estimated parameters and derived quantities, e.g., $M_{\rm SE}$, we confirm earlier findings of NLSy1s being AGN powered by rapidly accreting low-mass black holes and the steepening of the X-ray spectrum with increasing Eddington rate \citep[cf.][]{1996A&A...305...53B}. We conclude that with the advent of the ongoing and upcoming wide-field, multiwavelength sky surveys, e.g., VLASS, this new catalogue of NLSy1 galaxies will enable us to explore the physics of this enigmatic class of AGN in an unprecedented detail hence sowing the seeds for their future observations with the next-generation of giant telescopes.

The catalogue is made public at \url{https://www.ucm.es/blazars/seyfert} and also provided as a supplementary material of this article.

\begin{table*}
\caption{The \gm-ray detected sources present in the NLSy1 and BLSy1 catalogues. \label{tab:lat}
}
\begin{center}
\begin{tabular}{lllll}
\hline
4FGL Name & 4FGL Association & R.A.  (deg.)& Decl.  (deg.)&  Redshift \\
\hline
   &      & NLSy1 Galaxies & & \\
J0001.5+2113 &  TXS 2358+209              &  0.38487& 21.22672& 0.438\\
J0010.6+2043 &  TXS 0007+205              &  2.61976& 20.79716& 0.597\\
J0014.3$-$0500 &  WISE J001420.42$-$045928.7  & 3.58514& $-$4.99134& 0.790\\
J0038.7$-$0204 &  3C 17                     &  9.58555& $-$2.12790& 0.220\\
J0105.5+1912       & TXS 0103+189      & 16.48014 & 19.20778 & 0.879 \\
J0850.0+5108 &  SBS 0846+513              & 132.49155& 51.14139& 0.584\\
J0932.6+5306 &  S4 0929+53                &  143.17144& 53.10938& 0.597\\
J0933.5$-$0013  &  PMN J0933$-$0012 & 143.34596 & $-$0.18101 & 0.796 \\
J0948.9+0022 &  PMN J0948+0022            &  147.23882& 0.373753& 0.585\\
J0949.2+1749 &  TXS 0946+181              & 147.41565& 17.88037& 0.692\\
J0958.0+3222 &  3C 232                    &  149.58727& 32.40061& 0.530\\
J1127.8+3618  & MG2 J112758+3620 & 171.99528 & 36.34121 & 0.884 \\
J1305.3+5118 &  IERS B1303+515            & 196.34476& 51.27785& 0.787\\
J1331.0+3032 &  3C 286                    & 202.78456& 30.50913& 0.85 \\
J1443.1+4728 &  B3 1441+476               & 220.82733& 47.43241& 0.705\\
J1505.0+0326 &  PKS 1502+036              & 226.27698& 3.441895& 0.407\\
J1520.5+4209 &  B3 1518+423               & 230.16538& 42.18644& 0.485\\
J1639.2+4129 & MG4 J163918+4127    &249.81587 & 41.47603 & 0.690 \\
J1644.9+2620 &  MG2 J164443+2618          & 251.17722& 26.32035& 0.144\\
J2118.0+0019  & PMN J2118+0013  & 319.57249 & 0.22132 & 0.463 \\
J2118.8$-$0723 &  TXS 2116$-$077              & 319.72068& $-$7.54098& 0.260\\
\hline
   &      & BLSy1 Galaxies  & & \\
     J0059.2+0006 &  PKS 0056$-$00       &14.77297 &0.11434 &0.719\\
     J0112.1$-$0321 &  TXS 0110$-$037      &18.16305 &$-$3.47855 &0.772\\
     J0121.8+1147 &  PKS 0119+11       & 20.42330 &11.83067 &0.571\\
     J0136.8+2032 &  GB6 J0136+2028    &24.13413 &20.47153 &0.557\\
     J0203.7+3042 &  IVS B0200+30A     & 30.93447 &30.71052 &0.76 \\
     J0758.1+1134 &  TXS 0755+117      & 119.53189 &11.61279 &0.568\\
     J0814.4+2941 &  RX J0814.4+2941   &123.60795 &29.68770 &0.373\\
     J0827.8+5221 &  TXS 0824+524      & 126.97374 &52.29953 &0.338\\
     J0840.8+1317 &  3C 207            & 130.19828 &13.20654 &0.68 \\
     J0914.4+0249 &  PKS 0912+029      & 138.65796 &2.76645 &0.426\\
     J0924.0+2816 &  B2 0920+28        &140.96468 &28.25698 &0.744\\
     J0937.1+5008 &  GB6 J0937+5008    &144.30135 &50.14782 &0.275\\
     J0943.7+6137 &  GB6 B0940+6149    & 146.08517 &61.59726 &0.791\\
     J0945.2+5200 &  TXS 0941+522      & 146.21729 &52.04281  &0.5  \\
     J0956.7+2516  & OK 290                 & 149.20783  & 25.25446 & 0.708 \\
     J1014.8+2257   & OL 220                & 153.69611 & 23.02127   & 0.566 \\
     J1015.6+5553 &  TXS 1012+560      & 153.93508 &55.85017 &0.678\\
     J1036.2+2202 &  OL 256            & 159.13741 &22.05339 &0.595\\
     J1059.5+2057 &  MG2 J105938+2057  & 164.91266 &20.95608 &0.393\\
     J1102.6+5251 &  GB6 J1102+5249    & 165.70771 &52.83685 &0.69 \\
     J1102.9+3014 &  B2 1100+30B       & 165.80542 &30.24521 &0.384\\
     J1153.4+4931 &  4C +49.22         & 178.35194 &49.51910 &0.333\\
     J1159.5+2914 &  Ton 599           & 179.88265 &29.24551  &0.724\\
     J1209.8+1810 &  MG1 J120953+1809  & 182.46566 &18.16857 &0.845\\
     J1224.9+2122 &  4C +21.35         & 186.22691 &21.37955 &0.433\\
     J1257.8+3228 &  ON 393            & 194.48866 &32.49157 &0.805\\
     J1332.2+4722 &  B3 1330+476       & 203.18850 &47.37296 &0.669\\
     J1334.5+5634 &  TXS 1332+567      & 203.65616 &56.52997 &0.342\\
     J1350.8+3033 &  B2 1348+30B       & 207.71972 &30.58155 &0.712\\
     J1357.1+1921 &  4C +19.44         & 209.26846 &19.31872 &0.719\\
     J1422.5+3223 &  OQ 334            & 215.62657 &32.38623 &0.681\\
     J1430.5+3647   & TXS 1428+370  & 217.66911 & 36.81775 & 0.567 \\ 
     J1443.1+5201 &  3C 303            & 220.76150  &52.02701 &0.141\\
     J1524.2+1523 & MC3 1522+155 & 231.17338 & 15.35585 & 0.630 \\
     J1549.5+0236 &  PKS 1546+027      & 237.37265 &2.61697  &0.414\\
     J1604.6+5714 &  GB6 J1604+5714    & 241.15563 &57.24351 &0.721\\
     J1608.3+4012 & B2 1606+40  & 242.09232 & 40.20498 & 0.628 \\
     J1637.7+4717 &  4C +47.44         & 249.43805 &47.29273 &0.735\\
     J1642.9+3948 &  3C 345            & 250.74503 &39.81029 &0.593\\
     J1659.0+2627 &  4C +26.51         & 254.85061 &26.49360 &0.793\\
     J1706.9+4543 &  4C +45.34         &256.82397 &45.60292 &0.645\\
 \hline
\end{tabular}
\end{center}
\end{table*}

\begin{table*}
\contcaption{\label{tab:lat_2}}
\begin{center}
\begin{tabular}{lllll}
\hline
4FGL Name & 4FGL Association & R.A.  (deg.)& Decl.  (deg.)&  Redshift \\
\hline
   &      & BLSy1 Galaxies  & & \\
     J2157.0+1002 &  MC2 2154+100    & 329.30356 &10.24024 &0.761\\
     J2301.0$-$0158 &  PKS B2258$-$022     & 345.28322 &$-$1.96796 &0.777\\
     J2334.2+0736 & TXS 2331+073    & 353.55344 & 7.60764 & 0.401\\
     J2352.9+3031 &  MG3 J235254+3030  & 358.22795 &30.50603 &0.876\\  
 \hline
\end{tabular}
\end{center}
\end{table*}

\section*{Acknowledgements}
We are grateful to the referee for constructive criticism, which has helped improve the paper.  V.S.P. is grateful to Remington Sexton for a fruitful discussion on the use of \bd~software. A.D. is thankful for the support of the Ram{\'o}n y Cajal program from the Spanish MINECO, Proyecto PID2021-126536OA-I00 funded by MCIN / AEI / 10.13039/501100011033, and Proyecto PR44/21‐29915 funded by the Santander Bank and Universidad Complutense de Madrid.

Funding for the Sloan Digital Sky Survey IV has been provided by the Alfred P. Sloan Foundation, the U.S. Department of Energy Office of Science, and the Participating Institutions. SDSS-IV acknowledges support and resources from the Center for High-Performance Computing at the University of Utah. The SDSS web site is www.sdss.org.

SDSS-IV is managed by the Astrophysical Research Consortium for the Participating Institutions of the SDSS Collaboration including the Brazilian Participation Group, the Carnegie Institution for Science, Carnegie Mellon University, the Chilean Participation Group, the French Participation Group, Harvard-Smithsonian Center for Astrophysics, Instituto de Astrof\'isica de Canarias, The Johns Hopkins University, Kavli Institute for the Physics and Mathematics of the Universe (IPMU) / University of Tokyo, the Korean Participation Group, Lawrence Berkeley National Laboratory, Leibniz Institut f\"ur Astrophysik Potsdam (AIP),  Max-Planck-Institut f\"ur Astronomie (MPIA Heidelberg), Max-Planck-Institut f\"ur Astrophysik (MPA Garching), Max-Planck-Institut f\"ur Extraterrestrische Physik (MPE), National Astronomical Observatories of China, New Mexico State University, New York University, University of Notre Dame, Observat\'ario Nacional / MCTI, The Ohio State University, Pennsylvania State University, Shanghai Astronomical Observatory, United Kingdom Participation Group,Universidad Nacional Aut\'onoma de M\'exico, University of Arizona, University of Colorado Boulder, University of Oxford, University of Portsmouth, University of Utah, University of Virginia, University of Washington, University of Wisconsin, Vanderbilt University, and Yale University.

This publication makes use of data products from the Wide-field Infrared Survey Explorer, which is a joint project of the University of California, Los Angeles, and the Jet Propulsion Laboratory/California Institute of Technology, funded by the National Aeronautics and Space Administration.

This research has made use of the NASA/IPAC Extragalactic Database (NED), which is operated by the Jet Propulsion Laboratory, California Institute of Technology, under contract with the National Aeronautics and Space Administration. Part of this work is based on archival data, software or online services provided by the Space Science Data Center (SSDC). This research has made use of NASA's Astrophysics Data System Bibliographic Services.

This work made use of data supplied by the UK Swift Science Data Centre at the University of Leicester. This research has made use of data obtained from the Chandra Source Catalog, provided by the Chandra X-ray Center (CXC) as part of the Chandra Data Archive.
\section*{Data Availability}
All of the multi-wavelength data used in this article are publicly available at their respective data archives, e.g., SDSS-DR17 (\url{https://skyserver.sdss.org/dr17/}). The software used to analyze the SDSS spectra is publicly available (\url{https://github.com/remingtonsexton/BADASS3}).


\bibliographystyle{mnras}
\bibliography{Master} 

\appendix
\section{Details of the Catalogue Fits File}\label{sec:app}
We provide the full description of the fits catalogue below in Table~\ref{tab:nlsy1_cat}.

\begin{table*}
\caption{The format of the FITS catalogue. \label{tab:nlsy1_cat}}
\begin{center}
\begin{tabular}{llll}
\hline
Column  & Format & Units & Description \\
\hline
SDSS\_NAME & STRING &  & SDSS DR17 designation (J2000) \\
P-M-F              & STRING  &  & Spectroscopic Plate number, MJD, and fiber number \\
RA                    & DOUBLE & degree & Right ascension (J2000) \\
DEC                 & DOUBLE & degree & Declination (J2000) \\
Z                      & DOUBLE &      & Spectroscopic redshift \\
M\_B               & DOUBLE  &     & Absolute $B$-band magnitude \\
BR\_H\_ALPHA\_EW & DOUBLE & \AA  & Rest-frame equivalent-width of the broad \halpha~emission line \\
BR\_H\_ALPHA\_EW\_ERR & DOUBLE & \AA  & Uncertainty in rest-frame equivalent-width of the broad \halpha~emission line \\
BR\_H\_ALPHA\_FLUX & DOUBLE & [\ergflux]  & Flux of the broad \halpha~emission line on logarithmic scale \\
BR\_H\_ALPHA\_FLUX\_ERR & DOUBLE & [\ergflux] & Uncertainty in the logarithmic-scaled flux of the broad \halpha~emission line \\
BR\_H\_ALPHA\_FWHM & DOUBLE & \km  & FWHM of the broad \halpha~emission line \\
BR\_H\_ALPHA\_FWHM\_ERR & DOUBLE & \km  & Uncertainty in the FWHM of the broad \halpha~emission line \\
BR\_H\_ALPHA\_LUM & DOUBLE & [\lum]  & Luminosity of the broad \halpha~emission line on logarithmic scale \\
BR\_H\_ALPHA\_LUM\_ERR & DOUBLE & [\lum] & Uncertainty in the logarithmic-scaled luminosity of the broad \halpha~emission line \\
BR\_H\_ALPHA\_SNR & DOUBLE &   & S/N ratio of the broad \halpha~emission line \\
BR\_H\_BETA\_EW & DOUBLE & \AA  & Rest-frame equivalent-width of the broad \hbeta~emission line \\
BR\_H\_BETA\_EW\_ERR & DOUBLE & \AA  & Uncertainty in rest-frame equivalent-width of the broad \hbeta~emission line \\
BR\_H\_BETA\_FLUX & DOUBLE & [\ergflux]  & Flux of the broad \hbeta~emission line on logarithmic scale \\
BR\_H\_BETA\_FLUX\_ERR & DOUBLE & [\ergflux] & Uncertainty in the logarithmic-scaled flux of the broad \hbeta~emission line \\
BR\_H\_BETA\_FWHM & DOUBLE & \km  & FWHM of the broad \hbeta~emission line \\
BR\_H\_BETA\_FWHM\_ERR & DOUBLE & \km  & Uncertainty in the FWHM of the broad \hbeta~emission line \\
BR\_H\_BETA\_LUM & DOUBLE & [\lum]  & Luminosity of the broad \hbeta~emission line (\lum) on logarithmic scale \\
BR\_H\_BETA\_LUM\_ERR & DOUBLE & [\lum] & Uncertainty in the logarithmic-scaled luminosity of the broad \hbeta~emission line \\
BR\_H\_BETA\_SNR & DOUBLE &   & S/N ratio of the broad \hbeta~emission line \\
NA\_H\_ALPHA\_EW & DOUBLE & \AA  & Rest-frame equivalent-width of the narrow \halpha~emission line \\
NA\_H\_ALPHA\_EW\_ERR & DOUBLE & \AA  & Uncertainty in rest-frame equivalent-width of the narrow \halpha~emission line \\
NA\_H\_ALPHA\_FLUX & DOUBLE & [\ergflux]  & Flux of the narrow \halpha~emission line on logarithmic scale \\
NA\_H\_ALPHA\_FLUX\_ERR & DOUBLE & [\ergflux] & Uncertainty in the logarithmic-scaled flux of the narrow \halpha~emission line \\
NA\_H\_ALPHA\_FWHM & DOUBLE & \km  & FWHM of the narrow \halpha~emission line \\
NA\_H\_ALPHA\_FWHM\_ERR & DOUBLE & \km  & Uncertainty in the FWHM of the narrow \halpha~emission line \\
NA\_H\_ALPHA\_LUM & DOUBLE & [\lum]  & Luminosity of the narrow \halpha~emission line (\lum) on logarithmic scale \\
NA\_H\_ALPHA\_LUM\_ERR & DOUBLE & [\lum] & Uncertainty in the logarithmic-scaled luminosity of the narrow \halpha~emission line \\
NA\_H\_ALPHA\_SNR & DOUBLE &   & S/N ratio of the narrow \halpha~emission line \\
NA\_H\_BETA\_EW & DOUBLE & \AA  & Rest-frame equivalent-width of the narrow \hbeta~emission line \\
NA\_H\_BETA\_EW\_ERR & DOUBLE & \AA  & Uncertainty in rest-frame equivalent-width of the narrow \hbeta~emission line \\
NA\_H\_BETA\_FLUX & DOUBLE &  [\ergflux] & Flux of the narrow \hbeta~emission line on logarithmic scale \\
NA\_H\_BETA\_FLUX\_ERR & DOUBLE & [\ergflux] & Uncertainty in the logarithmic-scaled flux of the narrow \hbeta~emission line \\
NA\_H\_BETA\_FWHM & DOUBLE & \km  & FWHM of the narrow \hbeta~emission line \\
NA\_H\_BETA\_FWHM\_ERR & DOUBLE & \km  & Uncertainty in the FWHM of the narrow \hbeta~emission line \\
NA\_H\_BETA\_LUM & DOUBLE & [\lum]  & Luminosity of the narrow \hbeta~emission line (\lum) on logarithmic scale \\
NA\_H\_BETA\_LUM\_ERR & DOUBLE & [\lum] & Uncertainty in the logarithmic-scaled luminosity of the narrow \hbeta~emission line \\
NA\_H\_BETA\_SNR & DOUBLE &   & S/N ratio of the narrow \hbeta~emission line \\
NA\_OIII\_4960\_EW & DOUBLE & \AA  & Rest-frame equivalent-width of the narrow \OIIIa~emission line \\
NA\_OIII\_4960\_EW\_ERR & DOUBLE & \AA  & Uncertainty in rest-frame equivalent-width of the narrow \OIIIa~emission line \\
NA\_OIII\_4960\_FLUX & DOUBLE &  [\ergflux] & Flux of the narrow \OIIIa~emission line on logarithmic scale \\
NA\_OIII\_4960\_FLUX\_ERR & DOUBLE & [\ergflux] & Uncertainty in the logarithmic-scaled flux of the narrow \OIIIa~emission line \\
NA\_OIII\_4960\_FWHM & DOUBLE & \km  & FWHM of the narrow \OIIIa~emission line \\
NA\_OIII\_4960\_FWHM\_ERR & DOUBLE & \km  & Uncertainty in the FWHM of the narrow \OIIIa~emission line \\
NA\_OIII\_4960\_LUM & DOUBLE & [\lum]  & Luminosity of the narrow \OIIIa~emission line on logarithmic scale \\
NA\_OIII\_4960\_LUM\_ERR & DOUBLE & [\lum] & Uncertainty in the logarithmic-scaled luminosity of the narrow \OIIIa~emission line \\
NA\_OIII\_4960\_SNR & DOUBLE &   & S/N ratio of the narrow \OIIIa~emission line \\
NA\_OIII\_5007\_EW & DOUBLE & \AA  & Rest-frame equivalent-width of the narrow \OIIIb~emission line \\
NA\_OIII\_5007\_EW\_ERR & DOUBLE & \AA  & Uncertainty in rest-frame equivalent-width of the narrow \OIIIb~emission line \\
NA\_OIII\_5007\_FLUX & DOUBLE & [\ergflux]  & Flux of the narrow \OIIIb~emission line on logarithmic scale \\
NA\_OIII\_5007\_FLUX\_ERR & DOUBLE & [\ergflux] & Uncertainty in the logarithmic-scaled flux of the narrow \OIIIb~emission line \\
NA\_OIII\_5007\_FWHM & DOUBLE & \km  & FWHM of the narrow \OIIIb~emission line \\
NA\_OIII\_5007\_FWHM\_ERR & DOUBLE & \km  & Uncertainty in the FWHM of the narrow \OIIIb~emission line \\
NA\_OIII\_5007\_LUM & DOUBLE & [\lum]  & Luminosity of the narrow \OIIIb~emission line (\lum) on logarithmic scale \\
NA\_OIII\_5007\_LUM\_ERR & DOUBLE & [\lum] & Uncertainty in the logarithmic-scaled luminosity of the narrow \OIIIb~emission line \\
NA\_OIII\_5007\_SNR & DOUBLE &   & S/N ratio of the narrow \OIIIb~emission line \\
NA\_NII\_6549\_EW & DOUBLE & \AA  & Rest-frame equivalent-width of the narrow \NIIa~emission line \\
NA\_NII\_6549\_EW\_ERR & DOUBLE & \AA & Uncertainty in rest-frame equivalent-width of the narrow \NIIa~emission line \\
\hline
\end{tabular}
\end{center}
\end{table*}

\begin{table*}
\contcaption{\label{tab:nlsy1_c2}}
\begin{center}
\begin{tabular}{llll}
\hline
Column  & Format & Units & Description \\
\hline
NA\_NII\_6549\_FLUX & DOUBLE & [\ergflux]  & Flux of the narrow \NIIa~emission line on logarithmic scale \\
NA\_NII\_6549\_FLUX\_ERR & DOUBLE & [\ergflux] & Uncertainty in the logarithmic-scaled flux of the narrow \NIIa~emission line \\
NA\_NII\_6549\_FWHM & DOUBLE & \km  & FWHM of the narrow \NIIa~emission line \\
NA\_NII\_6549\_FWHM\_ERR & DOUBLE & \km  & Uncertainty in the FWHM of the narrow \NIIa~emission line \\
NA\_NII\_6549\_LUM & DOUBLE &  [\lum] & Luminosity of the narrow \NIIa~emission line (\lum) on logarithmic scale \\
NA\_NII\_6549\_LUM\_ERR & DOUBLE & [\lum] & Uncertainty in the logarithmic-scaled luminosity of the narrow \NIIa~emission line \\
NA\_NII\_6549\_SNR & DOUBLE &   & S/N ratio of the narrow \NIIa~emission line \\
NA\_NII\_6585\_EW & DOUBLE & \AA  & Rest-frame equivalent-width of the narrow \NIIb~emission line \\
NA\_NII\_6585\_EW\_ERR & DOUBLE & \AA  & Uncertainty in rest-frame equivalent-width of the narrow \NIIb~emission line \\
NA\_NII\_6585\_FLUX & DOUBLE & [\ergflux]  & Flux of the narrow \NIIb~emission line on logarithmic scale \\
NA\_NII\_6585\_FLUX\_ERR & DOUBLE & [\ergflux] & Uncertainty in the logarithmic-scaled flux of the narrow \NIIb~emission line \\
NA\_NII\_6585\_FWHM & DOUBLE & \km  & FWHM of the narrow \NIIb~emission line \\
NA\_NII\_6585\_FWHM\_ERR & DOUBLE & \km  & Uncertainty in the FWHM of the narrow \NIIb~emission line \\
NA\_NII\_6585\_LUM & DOUBLE & [\lum]  & Luminosity of the narrow \NIIb~emission line (\lum) on logarithmic scale \\
NA\_NII\_6585\_LUM\_ERR & DOUBLE & [\lum] & Uncertainty in the logarithmic-scaled luminosity of the narrow \NIIb~emission line \\
NA\_NII\_6585\_SNR & DOUBLE &   & S/N ratio of the narrow \NIIb~emission line \\
NA\_SII\_6718\_EW & DOUBLE & \AA  & Rest-frame equivalent-width of the narrow \SIIa~emission line \\
NA\_SII\_6718\_EW\_ERR & DOUBLE & \AA  & Uncertainty in rest-frame equivalent-width of the narrow \SIIa~emission line \\
NA\_SII\_6718\_FLUX & DOUBLE & [\ergflux]  & Flux of the narrow \SIIa~emission line (\ergflux) on logarithmic scale \\
NA\_SII\_6718\_FLUX\_ERR & DOUBLE & [\ergflux] & Uncertainty in the logarithmic-scaled flux of the narrow \SIIa~emission line \\
NA\_SII\_6718\_FWHM & DOUBLE & \km  & FWHM of the narrow \SIIa~emission line \\
NA\_SII\_6718\_FWHM\_ERR & DOUBLE & \km  & Uncertainty in the FWHM of the narrow \SIIa~emission line \\
NA\_SII\_6718\_LUM & DOUBLE & [\lum]  & Luminosity of the narrow \SIIa~emission line on logarithmic scale \\
NA\_SII\_6718\_LUM\_ERR & DOUBLE & [\lum] & Uncertainty in the logarithmic-scaled luminosity of the narrow \SIIa~emission line \\
NA\_SII\_6718\_SNR & DOUBLE &   & S/N ratio of the narrow \SIIa~emission line \\
NA\_SII\_6732\_EW & DOUBLE & \AA  & Rest-frame equivalent-width of the narrow \SIIb~emission line \\
NA\_SII\_6732\_EW\_ERR & DOUBLE & \AA  & Uncertainty in rest-frame equivalent-width of the narrow \SIIb~emission line \\
NA\_SII\_6732\_FLUX & DOUBLE & [\ergflux]  & Flux of the narrow \SIIb~emission line on logarithmic scale \\
NA\_SII\_6732\_FLUX\_ERR & DOUBLE & [\ergflux] & Uncertainty in the logarithmic-scaled flux of the narrow \SIIb~emission line \\
NA\_SII\_6732\_FWHM & DOUBLE & \km  & FWHM of the narrow \SIIb~emission line \\
NA\_SII\_6732\_FWHM\_ERR & DOUBLE & \km  & Uncertainty in the FWHM of the narrow \SIIb~emission line \\
NA\_SII\_6732\_LUM & DOUBLE &  [\lum] & Luminosity of the narrow \SIIb~emission line on logarithmic scale \\
NA\_SII\_6732\_LUM\_ERR & DOUBLE & [\lum] & Uncertainty in the logarithmic-scaled luminosity of the narrow \SIIb~emission line \\
NA\_SII\_6732\_SNR & DOUBLE &   & S/N ratio of the narrow \SIIb~emission line \\
L\_CONT\_AGN\_5100 & DOUBLE & [\lum] & Rest-frame continuum luminosity at 5100 \AA~(\lum) on logarithmic scale \\
L\_CONT\_AGN\_5100\_ERR & DOUBLE & [\lum] & Uncertainty in the logarithmic-scaled continuum luminosity at 5100 \AA \\
R4570                         & DOUBLE  &   & \FeII~strength \\
M\_BH & DOUBLE & [\msun] & Logarithmic-scaled single epoch black hole mass estimated using \hbeta~emission line parameters \\
M\_BH\_ERR & DOUBLE & [\msun] & Uncertainty in the logarithmic-scaled black hole mass estimated using \hbeta~emission line parameters \\
L\_BOL & DOUBLE & [\lum] & Bolometric luminosity on logarithmic scale \\
L\_BOL\_ERR & DOUBLE & [\lum] & Uncertainty in the logarithmic-scaled bolometric luminosity\\
R\_EDD & DOUBLE &  & Eddington ratio on logarithmic scale \\
R\_EDD\_ERR & DOUBLE &  & Uncertainty in the logarithmic-scaled Eddington ratio\\
RL & DOUBLE & & radio-loudness parameter on logarithmic scale \\
\hline
\end{tabular}
\end{center}
\end{table*}


\bsp	
\label{lastpage}
\end{document}